\documentclass[11pt]{elsarticle}
\usepackage{amssymb}
\usepackage{amsmath}
\usepackage{amsthm}
\usepackage{subfigure}
\usepackage{graphicx}

\biboptions{numbers,sort&compress}

\allowdisplaybreaks[4]  

\usepackage[dvipdfm,colorlinks,linkcolor=red,anchorcolor=blue,citecolor=blue]{hyperref}
\usepackage{booktabs}
\makeatletter

\newcommand{\Rmnum}[1]{\expandafter\@slowromancap\romannumeral #1@}
\makeatother
\usepackage{caption}

\usepackage{anysize}
\usepackage{setspace}
\marginsize{2.0cm}{1.5cm}{0.6cm}{0.6cm} 
\journal{Journal of XXX}

\begin{document}

\begin{frontmatter}

\title{The three--level coupled  Maxwell--Bloch equations: rogue waves, semirational rogue waves and W-shaped solitons}
\author{Xin Wang$^{{\rm a*}}$}\ead{wangxinlinzhou@163.com}
\author{Lei Wang$^{{\rm b}}$}
\author{Chong Liu$^{{\rm c,d}}$}
\cortext[cor1]{Corresponding author.}

\address{$^{{\rm a}}$College of Science, Zhongyuan University of Technology, Zhengzhou, 450007, China}
\address{$^{{\rm b}}$School of Mathematics and Physics, North China Electric Power University, Beijing, 102206, China}
\address{$^{{\rm c}}$School of Physics, Northwest University, Xi\rq an, 710069, China}
\address{$^{{\rm d}}$Shaanxi Key Laboratory for Theoretical Physics Frontiers, Xi'an 710069, China}

\begin{abstract}
In this paper the coupled  Maxwell--Bloch equations which describe the propagation
of two optical pulses in an optical medium with coherent three--level atoms are studied by Darboux transformation.
The general $n$th-order rogue wave solution involving two different choices of multiple roots for the
spectral characteristic equation and the multiparametric $n$th-order semirational solution
are both obtained in terms of Schur polynomials.
The explicit rogue wave solutions and semirational solutions from first to second order are provided.
In contrast to the known Peregrine soliton, dark and four-petaled structures,
some unusual patterns such as triple-hole, twisted-pair,
composite four-petaled and composite dark rogue waves are put forward.
Moreover, the interaction between dark-bright soliton and dark rogue wave and interaction
between breather and dark rogue wave are shown.
Further, the higher-order nonlinear superposition modes
which feature triple and quadruple temporal-spatial distributions are presented.
Finally, the state transition between rogue wave and
W-shaped soliton is found where the modulation instability growth rate tends to
zero under the low perturbation frequency. Particularly, the dark and double-peak W-shaped solitons are examined.
\end{abstract}

\begin{keyword}
The three--level coupled Maxwell--Bloch equations; Rogue waves; Semirational rogue waves;
W-shaped solitons; Darboux transformation; Modulation instability\\
\end{keyword}
\end{frontmatter}

\section{Introduction}  


In the past few decades, the field of fiber optical communication has been a rapid development
in theoretical and experimental studies. Particularly, the concept of soliton has been applied to
describe intense electromagnetic beams and ultrashort pulse propagation in diverse medium.
Dissipation and dispersion in the medium are
the main obstacle for signals transmitting through optical fibers. In order to counter this,
transmission of picosecond optical pulses simulated by the
soliton solution of the celebrated nonlinear Schr\"{o}dinger (NLS) equation,
in which the group velocity dispersion can be balanced by the phase self-modulation, was investigated
by Hasegawa and Tappert forty years ago \cite{01}. Another type of optical soliton is based on the
self-induced transparency (SIT) effect in coherent medium associated with an incident electric field,
to which the medium is wholly transparent and the propagation is lossless.
This phenomenon was first discovered by McCall and Hahn when studying the optical soliton
in a two-level resonant system in 1967 \cite{02}.

The SIT phenomenon in a two-level medium can be governed by the Maxwell-Bloch equations \cite{03}.
Some classical methods in the soliton theory
such as the inverse scattering transform \cite{04}, the Darboux transformation \cite{05} and
the Riemann-Hilbert approach \cite{06} have been actively utilized to solve the MB equations
with inhomogeneous broadening and sharp-line limit case.
Particularly, a variety of special solutions such as rogue wave solutions
for the Maxwell-Bloch equations have been exactly proposed to describe the propagation
of optical pulses in a two-level optical medium \cite{07,08,09}.
More recently, a particular class of solutions for the reduced Maxwell-Bloch equations,
the generalized Maxwell-Bloch equations and the relevant \lq\lq AB\rq\rq~system have been extensively
investigated by many authors \cite{10,11,12,13,14}.


It is also important to consider the pulse propagation in a three-level system,
which can be governed by the following coupled Maxwell-Bloch (CMB) equations:
\begin{subequations}\label{02}
\begin{align}
&E_{1t}=\langle p_{1}\rangle,\label{e21}\\
&E_{2t}=\langle p_{2}\rangle,\label{e22} \\
&p_{1x}+2{\rm i}\omega p_{1}=\frac{1}{2}\left(M_{11}E_1+M_{21}E_2+NE_1\right),\label{e23}\\
&p_{2x}+2{\rm i}\omega p_{2}=\frac{1}{2}\left(M_{12}E_1+M_{22}E_2+NE_2\right),\label{e24}\\
&N_x=-\frac{1}{2}\left(p_1 E_1^{*}+p_1^{*}E_1\right)-\frac{1}{2}\left(p_2 E_2^{*}+p_2^{*}E_2\right),\label{e25}\\
&M_{11x}=-\frac{1}{2}(p_1 E_1^{*}+p_1^{*}E_1),\label{e26}\\
&M_{12x}=-\frac{1}{2}(p_1^{*} E_2+p_{2}E_1^{*}),\label{e27}\\
&M_{21x}=-\frac{1}{2}(p_1 E_2^{*}+p_2^{*}E_1),\label{e28}\\
&M_{22x}=-\frac{1}{2}(p_2 E_2^{*}+p_2^{*}E_2),\label{e29}
\end{align}
\end{subequations}
where the brackets $\langle \rangle$  denotes the averaging over $\omega$ with
given distribution function $g(\omega)$, namely,
$$
\langle p_{j}\rangle=\int_{-\infty}^{+\infty}p_{j}(x,t;\omega)g(\omega)d\omega,\ j=1,2.
$$
Here the sign $*$ stands for the complex conjugate, $E_1$ and $E_2$ represent two
propagating electric fields, $N$, $p_1$, $p_2$ and $M_{ij}$ ($i,j=1,2$) are elements of the
density matrix of atomic subsystem
$$
\widetilde{p}=\left(
              \begin{array}{ccc}
                -N & p_{1} & p_{2} \\
                p_{1}^{*} & M_{11} & M_{12} \\
                p_{2}^{*} & M_{21} & M_{22} \\
              \end{array}
            \right),$$
with $M_{21}=M_{12}^{*}$.
Eqs. (\ref{e23})-(\ref{e29}) can be casted into the following matrix form, as
\begin{equation}\label{}
\widetilde{p}_x=\dfrac{1}{2}\left[-2{\rm i}\omega\sigma_3+\widetilde{E},\widetilde{p}\right],
\end{equation}
where $[\cdot,\cdot]$ represents the matrix commutator, and
$$
\sigma_3=\left(
           \begin{array}{ccc}
             1 & 0 & 0 \\
             0 & -1 & 0 \\
             0 & 0 & -1 \\
           \end{array}
         \right),\
\widetilde{E}=\left(
              \begin{array}{ccc}
                0 & E_{1} & E_{2} \\
               -E_{1}^{*} & 0 & 0 \\
                -E_{2}^{*} & 0 & 0 \\
              \end{array}
            \right).$$
The CMB system (\ref{02}) has been an active research subject
during the past few decades due to its close connection with the SIT \cite{15} and
electromagnetically induced transparency  \cite{16}.
These problems are usually related to the ${\rm \Lambda}$,
V and the \lq\lq cascade\rq\rq~configurations of three-level medias \cite{15}.
The complete integrability, soliton
solution and breather solution for the CMB equations
have been studied by Wadati, Ablowitz, Shin and other
authors \cite{17,18,19,20}. Here, it should be pointed out that
the integral associated with the distribution function $g(\omega)$ accounts for the
inhomogeneous broadening effect, while for the case of sharp-line limit
whence $g(\omega)$ tends to a Dirac-delta function at the resonant frequency,
the bracket could be eliminated, viz., $\langle p_j\rangle\equiv p_j$, ($j=1,2$).
In this paper, we consider the simple sharp-line limit for the CMB equations.


After several years of rapid development, rogue waves have nowadays gone deep into
almost every corner of modern science, ranging from oceanography \cite{21} and optics \cite{22} to
Bose-Einstein condensates \cite{23} and even finance \cite{24}.
A single rogue wave can be described by the famous Peregrine soliton
of the NLS equation \cite{25}, and further the superposition of a certain mount of rogue waves
are simulated by higher-order rational solutions of the NLS equation \cite{26,27,28,29}.
Generally speaking, a wave that has a peak three times larger than the background amplitude,
localizes in both space and time, and appears to follow $L$-shaped statistics
can be classified into rogue wave category \cite{22,25,26,27,28,29,YJK}. Nevertheless, for a comprehensive
understanding of the rogue wave phenomenon, one should concentrate on not only the single-wave
NLS model but also the coupled systems with more than one equation.
In this direction, rogue wave solutions of the coupled NLS equations \cite{30},
the coupled Hirota equations \cite{31,32}, the three-wave
resonant interaction equations \cite{33,34} and other coupled systems \cite{35,CJC}
have been extensively studied.
As a matter of fact, apart from the Peregrine soliton type,
some new rogue wave patterns, such as dark \cite{35}, four-petaled \cite{36}, composite \cite{30} and
semirational rogue waves \cite{37,38} can exist in the coupled systems
with different modes, frequencies or polarizations.

In this paper, we focus on rogue waves, semirational rogue waves and W-shaped solitons
in the CMB equations by making use of the Darboux transformation (DT)
method and the modulation instability analysis \cite{39,40,41}.
Our paper can be organized as follows. In section 2, we present a Lax pair associated with
its $n$-fold DT based on the standard procedure
for the Ablowitz-Kaup-Newell-Segur (AKNS) problem.
In section 3, by taking into account of the double-root and triple-root
distributions for the spectral characteristic equation,
we derive the general $n$th-order rogue wave solution in terms of Schur polynomials
under one specific plane-wave background.
We show some rare rogue wave patterns
such as triple-hole rogue wave and twisted-pair rogue wave.
In section 4,
we obtain the multiparametric $n$th-order semirational rogue wave solution in terms of Schur polynomials
under another plane-wave background.
The extraordinary interaction between dark-bright soliton and dark rogue wave
and that between breather and dark rogue wave are exhibited for illustration.
In section 5, we apply the standard linearized stability analysis
to establish the analytical parametric condition of the sate transition
between rogue wave and W-shaped soliton.
The dark and double-peak W-shaped solitons are presented as examples.
Finally, we give our conclusion in section 6.


\section{Lax pair and Darboux transformation}

It is known that the CMB system (\ref{02}) admits the following $3\times 3$ linear
eigenvalue problem \cite{18}:
\begin{subequations}\label{LP}
\begin{align}
&\Phi_{x}=U\Phi,\ U=\dfrac{1}{2}(-{\rm i}\lambda\sigma_{3}+\widetilde{E}),\label{IST}\\
&\Phi_{t}=V\Phi,\ V=\dfrac{1}{2{\rm i}(\lambda-2\omega)}\widetilde{p},\label{IST2}
\end{align}
\end{subequations}
where $\lambda$ is the spectral parameter and $\Phi=(\psi,\varphi,\chi)^{T}$ is the complex eigenfunction.
One can directly recover the CMB system (\ref{02}) through the compatibility condition $U_t-V_x+[U,V]=0$.

After that, we establish the standard DT for System (\ref{LP}) by using the conventional
gauge transformation method \cite{05}. Suppose $\Phi_{j}=(\psi_{j},\varphi_{j},\chi_{j})^{T}$ ($1\leq j\leq n$)
are $n$ distinct solutions of Eqs. (\ref{IST}) and (\ref{IST2}) at $\lambda=\lambda_j$,
then the $n$-fold DT for System (\ref{LP}) takes the form
\begin{equation}\label{nDT}
\Phi[n]=T\Phi, T=I+\sum_{j=1}^{n}\dfrac{T_{j}}{\lambda-\lambda_{j}^{*}}=I-JH^{-1}D^{-1}J^{\dag},
\end{equation}
where
$$\begin{array}{l}
J=(\Phi_1,\Phi_2,\cdots,\Phi_n),\\
D={\rm diag}(\lambda-\lambda_1^{*},\lambda-\lambda_2^{*},\cdots,\lambda-\lambda_n^{*}),\\
H=(H_{ij})_{n\times n}=\left(\dfrac{\Phi_{i}^{\dag}\Phi_{j}}{\lambda_{j}-\lambda_{i}^{*}}\right).
\end{array}$$
Here the symbol $\dag$ denotes the complex conjugate transpose.
It follows that the $n$-fold DT keeps System (\ref{LP}) invariant, which implies
$$\begin{array}{l}
\Phi[n]_{x}=U[n]\Phi[n],\ U[n]=\dfrac{1}{2}(-{\rm i}\lambda\sigma_{3}+\widetilde{E}[n]),\\
\Phi[n]_{t}=V[n]\Phi[n],\ V[n]=\dfrac{1}{2{\rm i}(\lambda-2\omega)}\widetilde{p}[n],
\end{array}$$
such that
$$
\widetilde{E}[n]=\left(
              \begin{array}{ccc}
                0 & E_{1}[n] & E_{2}[n] \\
               -E_{1}[n]^{*} & 0 & 0 \\
                -E_{2}[n]^{*} & 0 & 0 \\
              \end{array}
            \right),\ \widetilde{p}=\left(
              \begin{array}{ccc}
                -N[n] & p_{1}[n] & p_{2}[n] \\
                p_{1}[n]^{*} & M_{11}[n] & M_{12}[n] \\
                p_{2}[n]^{*} & M_{21}[n] & M_{22}[n] \\
              \end{array}
            \right),
$$
where
\begin{subequations}\label{07}
\begin{align}
&E_{1}[n]=E_{1}+2{\rm i}\dfrac{1}{|H|}\left|
                                \begin{array}{cc}
                                  H & Y^{\dag} \\
                                  X & 0 \\
                                \end{array}
                              \right|,\label{71}\\
&E_{2}[n]=E_{2}+2{\rm i}\dfrac{1}{|H|}\left|
                                \begin{array}{cc}
                                  H & Z^{\dag} \\
                                  X & 0 \\
                                \end{array}
                              \right|,\\
&N[n]=N-2{\rm i}\dfrac{\partial}{\partial t}\dfrac{1}{|H|}\left|
                                \begin{array}{cc}
                                  H & X^{\dag} \\
                                  X & 0 \\
                                \end{array}
                              \right|,\\
&p_{1}[n]=p_{1}+2{\rm i}\dfrac{\partial}{\partial t}\dfrac{1}{|H|}\left|
                                \begin{array}{cc}
                                  H & Y^{\dag} \\
                                  X & 0 \\
                                \end{array}
                              \right|,\\
&p_{2}[n]=p_{2}+2{\rm i}\dfrac{\partial}{\partial t}\dfrac{1}{|H|}\left|
                                \begin{array}{cc}
                                  H & Z^{\dag} \\
                                  X & 0 \\
                                \end{array}
                              \right|,\\
&M_{11}[n]=M_{11}+2{\rm i}\dfrac{\partial}{\partial t}\dfrac{1}{|H|}\left|
                                \begin{array}{cc}
                                  H & Y^{\dag} \\
                                  Y & 0 \\
                                \end{array}
                               \right|,\\
&M_{12}[n]=M_{12}+2{\rm i}\dfrac{\partial}{\partial t}\dfrac{1}{|H|}\left|
                                \begin{array}{cc}
                                  H & Z^{\dag} \\
                                  Y & 0 \\
                                \end{array}
                              \right|,\\
&M_{21}[n]=M_{21}+2{\rm i}\dfrac{\partial}{\partial t}\dfrac{1}{|H|}\left|
                                \begin{array}{cc}
                                  H & Y^{\dag} \\
                                  Z & 0 \\
                                \end{array}
                              \right|,\\
&M_{22}[n]=M_{22}+2{\rm i}\dfrac{\partial}{\partial t}\dfrac{1}{|H|}\left|
                                \begin{array}{cc}
                                  H & Z^{\dag} \\
                                  Z & 0 \\
                                \end{array}
                              \right|,\label{79}
\end{align}
\end{subequations}
with
$$
X=(\psi_{1},\psi_{2},\cdots,\psi_{n}),\ Y=(\varphi_{1},\varphi_{2},\cdots,\varphi_{n}),\
Z=(\chi_{1},\chi_{2},\cdots,\chi_{n}).
$$
The formulas (\ref{71})-(\ref{79}) together with the series expansion technique
will be employed to derive the explicit rogue wave and semirational solutions for the CMB equations.

\section{Rogue wave solutions }

In this section, we begin with a plane-wave solution of the CMB equations
\begin{equation}\label{08}
\begin{array}{l}
E_{j}[0]=c_{j}{\rm e}^{{\rm i}\theta_{j}},\
p_{j}[0]={\rm i}c_{j}b_{j}{\rm e}^{{\rm i}\theta_{j}}, \ M_{jj}[0]=m_{j},j=1,2,\\
N[0]=N_{0},\ M_{12}[0]=M_{21}[0]^{*}=-\dfrac{c_{1}c_{2}\sigma}{2\delta}{\rm e}^{{\rm i}(\theta_{2}-\theta_{1})},
\end{array}
\end{equation}
where $\theta_{j}=a_{j}x+b_{j}t$, $\delta=a_1-a_2$, $\sigma=b_1-b_2$,
$a_j$, $b_j$, $c_j$ stand for the
frequency, wavenumber and background of the electric field $E_j$, respectively,
$N_0$ and $m_j$ are real constants and yield the relation
\begin{align}
&m_{1}=-\dfrac{1}{2\delta}[2(2a_{1}b_{1}+4b_{1}\omega+N_{0})\delta-c_{2}^2\sigma],\nonumber\\
&m_{2}=-\dfrac{1}{2\delta}[2(2a_{2}b_{2}+4b_{2}\omega+N_{0})\delta-c_{1}^2\sigma].\nonumber
\end{align}
Substitution of the above plane-wave solution into Eqs. (\ref{IST}) and (\ref{IST2}) can
lead to the fundamental solution
\begin{equation}\label{phieq}
\Phi=G\left(
       \begin{array}{ccc}
         1 & 1 & 1 \\
         \dfrac{{\rm i}c_{1}}{2\xi_{1}-\lambda-2a_{1}} & \dfrac{{\rm i}c_{1}}{2\xi_{2}-\lambda-2a_{1}} & \dfrac{{\rm i}c_{1}}{2\xi_{3}-\lambda-2a_{1}} \\
         \dfrac{{\rm i}c_{2}}{2\xi_{1}-\lambda-2a_{2}} & \dfrac{{\rm i}c_{2}}{2\xi_{2}-\lambda-2a_{2}} & \dfrac{{\rm i}c_{2}}{2\xi_{3}-\lambda-2a_{2}} \\
       \end{array}
     \right)\left(
              \begin{array}{c}
                {\rm e}^{A_{1}} \\
                {\rm e}^{A_{2}} \\
                {\rm e}^{A_{3}} \\
              \end{array}
            \right),
\end{equation}
where $G={\rm diag}(1,{\rm e}^{-{\rm i}\theta_{1}},{\rm e}^{-{\rm i}\theta_{2}})$, $\xi_{j}$ ($j=1,2,3$)
satisfy the cubic spectral characteristic equation
\begin{equation}\label{09}
\xi^3-\left(\dfrac{1}{2}\lambda+\varsigma\right)\xi^2-\left(\dfrac{1}{4}\lambda^2
+\dfrac{1}{4}\varrho-a_{1}a_{2}\right)\xi+\dfrac{1}{8}\lambda^3+\dfrac{1}{4}\varsigma\lambda^2
+\dfrac{1}{8}(\varrho+4a_{1}a_{2})\lambda+\dfrac{1}{4}c_1^2a_2+\dfrac{1}{4}c_2^2a_1=0,
\end{equation}
with $\varsigma=a_1+a_2$, $\varrho=c_1^2+c_2^2$, and
\begin{equation}\label{Aj}
A_j={\rm i}\left(\xi_jx+\dfrac{[\sigma\xi_{j}^2+(a_1b_2-a_2b_1)\xi_{j}]}{(\lambda-2\omega)\delta}t\right),\ j=1,2,3.
\end{equation}

As known in Refs. \cite{30}, the generalized DT which can be viewed as a special limit of the $n$-fold DT
is one of the most effective tools to construct higher-order rogue wave solutions.
To this end, one should select a special spectral parameter $\lambda_1$ and a
multiple root of the spectral characteristic equation for given parameters in the plane-wave solution.
In what follows, we will apply the generalized DT to derive the general
$n$th-order rogue wave solution
by considering the double-root and triple-root cases of Eq. (\ref{09}).

\subsection{Series expansions of double-root case}

The discriminant of the cubic equation (\ref{09}) can be written as
\begin{equation}\label{10}
\Delta=-4\rho^3+27\upsilon^2,
\end{equation}
where
\begin{align}
&\rho=\dfrac{1}{3}\lambda^2+\dfrac{1}{3}\varsigma\lambda+\dfrac{1}{4}\varrho+\dfrac{1}{3}(a_1^2-a_1a_2+a_2^2),
\nonumber\\
&\upsilon=\dfrac{2}{27}\lambda^3+\dfrac{1}{9}\varsigma\lambda^2
+\dfrac{1}{3}\left(\dfrac{1}{4}\varrho-\dfrac{1}{3}a_1^2
+\dfrac{4}{3}a_1a_2-\dfrac{1}{3}a_2^2\right)\lambda+\dfrac{1}{6}(c_1^2a_2+c_2^2a_1)\nonumber\\
&~~-\dfrac{1}{12}(c_1^2a_1+c_2^2a_2)-\dfrac{2}{27}(a_1^2-a_1a_2+a_2^2)\varsigma
+\dfrac{1}{9}a_1a_2\varsigma.\nonumber
\end{align}
One can prove Eq. (\ref{09}) possesses a double root when $\Delta=0$ for
the nonzero $\rho$ and $\upsilon$.

At this point, we assume that a special spectral parameter $\lambda=\lambda_1$ fulfills Eq. (\ref{10})
under the given plane-wave solution parameters and corresponds to a double root of Eq. (\ref{09}),
which, we denote $\xi_2=\xi_1\neq\xi_3$. Then a perturbation of the spectral parameter
\begin{equation}\label{lam}
\lambda_1(\epsilon)=\lambda_1+\ell\epsilon^2,
\end{equation}
where $\epsilon$ is a small complex parameter and
$$
\ell=\dfrac{2(\lambda_1+2\varsigma-6\xi_1)}{4 \xi_1^2-4(\lambda_1+\varsigma) \xi_1
+\lambda_1^2+2\varsigma\lambda_1 +4 a_1 a_2},
$$
can give rise to the expansion
\begin{equation}
\xi_{1}(\epsilon)=\sum_{j=0}^{\infty}\xi_1^{[j]}\epsilon^j,
\end{equation}
where
$$
\begin{array}{l}
\xi_{1}^{[0]}=\xi_{1},\ \xi_{1}^{[1]}=1,\
\xi_{1}^{[2]}=-\dfrac{2 \ell \xi_{1}+\ell \lambda_1-2}{2(\lambda_1+2\varsigma-6 \xi_1)},\\
\xi_1^{[3]}=\dfrac{1}{8(\lambda_1+2\varsigma-6 \xi_1)}
\left[\left(24(\xi_{1}^{[2]})^2-8 \ell \xi_{1}^{[2]}-2 \ell^2\right) \xi_1\right.
\\ ~~~~\left.-\left(4\lambda_1+8\varsigma\right)(\xi_{1}^{[2]})^2
-\left(4 \ell\lambda_1-24\right)\xi_{1}^{[2]}+2 \varsigma \ell^2
+3 \ell^2 \lambda_1-4 \ell\right],\\
\xi_1^{[4]}=-\dfrac{1}{4(\lambda_1+2\varsigma-6 \xi_1)}
\left[\left(4 \ell \xi_{1}^{[3]}-24 \xi_{1}^{[2]} \xi_{1}^{[3]}\right) \xi_1-12 (\xi_{1}^{[2]})^2\right.\\
~~~~\left.+\left((4 \lambda_1+8\varsigma) \xi_{1}^{[3]}+4 \ell\right)\xi_{1}^{[2]}+(2 \ell \lambda_1-12) \xi_{1}^{[3]}+\ell^2\right],\\
\xi_1^{[5]}=-\dfrac{1}{8(\lambda_1+2\varsigma-6 \xi_1)}
\left[\left(8 \ell \xi_1^{[4]}-48 \xi_1^{[2]} \xi_1^{[4]}-24 (\xi_1^{[3]})^2\right) \xi_1\right.\\
~~~~-8(\xi_1^{[2]})^3+4 \ell (\xi_1^{[2]})^2-\left(48 \xi_1^{[3]}-(8\lambda_1
+16\varsigma) \xi_1^{[4]}-2 \ell^2\right) \xi_1^{[2]}\\
~~~~\left.+(4 \lambda_1+8\varsigma) (\xi_1^{[3]})^2+8 \ell \xi_1^{[3]}+(4 \ell \lambda_1-24) \xi_1^{[4]}-\ell^3\right],
\end{array}
$$
and $\xi_{1}^{[j]}$ ($j\geq 6$) satisfy
$$
\begin{array}{l}
\xi_1^{[j]}=\left[\displaystyle\sum_{m+n+k=j+1}^{0\leq m,n,k\leq j-1}\xi_1^{[m]}\xi_1^{[n]}\xi_1^{[k]}-\left(\dfrac{1}{2}\lambda_1+\varsigma\right)\sum_{m+n=j+1}^{0\leq m,n\leq j-1}\xi_1^{[m]}\xi_1^{[n]}-\dfrac{1}{2}\ell\sum_{m+n=j-1}\xi_1^{[m]}\xi_1^{[n]}\right.\\
~~~~~~~\left.-\dfrac{1}{2}\ell\lambda_1\xi_1^{[j-1]}-\dfrac{1}{4}\ell^2\xi_1^{[j-3]}\right]
/(\lambda_1+2\varsigma-6 \xi_1).
\end{array}$$
We then have
\begin{equation}\label{13}
A_1(\epsilon)=\sum_{j=0}^{\infty}A_1^{[j]}\epsilon^j,
\end{equation}
where
$$
A_1^{[j]}={\rm i}\left(\xi_1^{[j]}x+\dfrac{1}{\delta}\left[\sum_{k=0}^{[\frac{j}{2}]}\sum_{l=0}^{j-2k}
\dfrac{(-\ell)^k\sigma}{(\lambda_1-2\omega)^{k+1}}
\xi_1^{[l]}\xi_1^{[j-2k-l]}
+\sum_{k=0}^{[\frac{j}{2}]}\dfrac{(-\ell)^k(a_1b_2-a_2b_1)}{(\lambda_1-2\omega)^{k+1}}\xi_1^{[j-2k]} \right]t\right).
$$
Here $[m]$ that is on top of the summation sign denotes the integer part of a real number $m$.

Furthermore, based on the Schur polynomials $S^{[j]}(\mathbf{x})$, which are defined by
$$
\sum_{j=0}^{\infty}S^{[j]}(\mathbf{x})\epsilon^j=\exp\left(\sum_{j=1}^{\infty}x_j\epsilon^j\right),
$$
where $\mathbf{x}=(x_1,x_2,\cdots)$,
$$
S^{[j]}=\sum_{\sum_{k=0}^{m}kl_k=j}\dfrac{(x_1)^{l_1}(x_2)^{l_2}
\cdots(x_m)^{l_m}}{l_1!l_2!\cdots l_m!},
$$
and specially,
$$
S^{[0]}=1,\ S^{[1]}=x_1,\ S^{[2]}=\dfrac{1}{2}x_1^2+x_2,\
S^{[3]}=x_3+x_1x_2+\dfrac{1}{6}x_1^3,
$$
one obtains
\begin{equation}
\exp\left(\sum_{j=1}^{\infty}A_1^{[j]}\epsilon^j\right)=
\sum_{j=0}^{\infty}S^{[j]}({\bf A})\epsilon^j,
\end{equation}
where $\mathbf{A}=(A_1^{[1]},A_1^{[2]},\cdots)$.

On the other hand, using the expansion
\begin{equation}
\dfrac{1}{2\xi_{1}(\epsilon)-\lambda_1(\epsilon)-2a_{l}}=\sum_{j=0}^{\infty}\mu_{l}^{[j]}\epsilon^j,\ l=1,2,
\end{equation}
where
$$
\begin{array}{l}
\mu_{l}^{[0]}=\dfrac{-1}{-\lambda_1-2a_{l}+2\xi_1},\ \mu_{l}^{[1]}=\dfrac{-2}{(-\lambda_1-2a_{l}+2\xi_1)^2},\\
\mu_{l}^{[k]}=\displaystyle\dfrac{-1}{-\lambda_1-2a_{l}+2\xi_1}\left[\sum_{m=1}^{k}2\mu_{l}^{[k-m]}\xi_1^{[m]}
-\ell\mu_{l}^{[k-2]}  \right],\ k\geq 2,
\end{array}$$
we can define
\begin{equation}
\Upsilon(\epsilon)=\sum_{j=0}^{\infty}\Upsilon^{[j]}\epsilon^j,
\end{equation}
where
$$
\Upsilon^{[j]}=\left(
                         \begin{array}{c}
                           S^{[j]}(\mathbf{A}) \\
                           {\rm i}c_1\displaystyle\sum^{j}_{k=0}\mu_{1}^{[j-k]}S^{[k]}(\mathbf{A}) \\
                           {\rm i}c_2\displaystyle\sum^{j}_{k=0}\mu_{2}^{[j-k]}S^{[k]}(\mathbf{A}) \\
                         \end{array}
                       \right),
$$

At this stage, we introduce
\begin{equation}\label{19}
\Psi_1(\epsilon)=\Gamma_1
\left[\dfrac{\Upsilon(\epsilon)+\Upsilon(-\epsilon)}{2}\right]
+\Gamma_2\left[\dfrac{\Upsilon(\epsilon)-\Upsilon(-\epsilon)}{2\epsilon}\right],
\end{equation}
where
$\Gamma_1=\sum_{k=0}^{\infty}\gamma_{2k}\epsilon^{2k},\ \Gamma_2=\sum_{k=0}^{\infty}\gamma_{2k+1}\epsilon^{2k}$, $\gamma_j$ ($j\geq0$) are complex constants. Then $\Psi_1$ can be expanded as the series
with terms in powers of $\epsilon^2$, that is
\begin{equation}\label{psi1}
\Psi_1(\epsilon)=\sum_{j=0}^{\infty}\Psi_1^{[j]}\epsilon^{2j},
\end{equation}
where
$$
\Psi_1^{[j]}=\sum_{k=0}^{j}\left[\gamma_{2k}\Upsilon^{[2(j-k)]}+\gamma_{2k+1}
\Upsilon^{[2(j-k)+1]}\right].
$$
It should be pointed out that $\Psi_1^{[j]}=(\psi_1^{[j]},\varphi_1^{[j]},\chi_1^{[j]})^T$ and
$G\Psi_1$ is the solution of System (\ref{LP}) under the plane-wave solution (\ref{08})
and the spectral parameter (\ref{lam}).

Returning to the $n$-fold DT (\ref{nDT}) we notice that Eqs. (\ref{lam}) and (\ref{psi1}) yield
\begin{equation}\label{21}
\dfrac{1}{\lambda_1(\epsilon)-\lambda_1(\epsilon)^{*}}
=\sum_{j=0}^{\infty}\sum_{m=0}^{j}\left(
\begin{array}{c}
j \\
m \\
\end{array}
\right)\dfrac{\ell^{*(j-m)}(-\ell)^{m}}{(\lambda_1-\lambda_1^{*})^{j+1}}\epsilon^{*2(j-m)}\epsilon^{2m},
\end{equation}
where
$$
\left(
\begin{array}{c}
j \\
m \\
\end{array}
\right)=\dfrac{j(j-1)\cdots (j-m+1)}{m!},
$$
and
\begin{equation}\label{22}
\Psi_1(\epsilon)^{\dag}\Psi_1(\epsilon)
=\sum_{j=0}^{\infty}\sum_{m=0}^{j}\Psi_1^{[j-m]\dag}\Psi_1^{[m]}\epsilon^{*2(j-m)}\epsilon^{2m}.
\end{equation}
Now we arrive at
$$
\begin{array}{l}
\dfrac{\Psi_1(\epsilon)^{\dag}\Psi_1(\epsilon)}{\lambda_1(\epsilon)-\lambda_1(\epsilon)^{*}}=
\displaystyle\sum_{j=0}^{\infty}\sum_{l=0}^{j}\left[\sum_{m=0}^{l}
\left(
\begin{array}{c}
l \\
m \\
\end{array}
\right)\dfrac{\ell^{*(l-m)}(-\ell)^{m}}{(\lambda_1-\lambda_1^{*})^{l+1}}\epsilon^{*2(l-m)}\epsilon^{2m}\right]
\left[\sum_{n=0}^{j-l}\Psi_1^{[j-l-n]\dag}\Psi_1^{[n]}\epsilon^{*2(j-l-n)}\epsilon^{2n}\right]\\
\hspace{2.7cm}=\displaystyle\sum_{j=0}^{\infty}\sum_{s=0}^{j}\left[\sum_{l=0}^{j}
\sum_{m+n=s}^{\mbox{\tiny$\begin{array}{c}
0\leq m\leq l,\\
0\leq n\leq j-l\\
\end{array}$}}
\left(
\begin{array}{c}
l \\
m \\
\end{array}
\right)\dfrac{\ell^{*(l-m)}(-\ell)^{m}}{(\lambda_1-\lambda_1^{*})^{l+1}}
\Psi_1^{[j-l-n]\dag}\Psi_1^{[n]}\right]\epsilon^{*2(j-s)}\epsilon^{2s}.
\end{array}$$
Then it turns out to be
\begin{equation}\label{23}
\dfrac{\Psi_1(\epsilon)^{\dag}\Psi_1(\epsilon)}{\lambda_1(\epsilon)-\lambda_1(\epsilon)^{*}}=
\sum_{r,t=1}^{\infty}H_1^{[r,t]}\epsilon^{*2(r-1)}\epsilon^{2(t-1)},
\end{equation}
where
$$
H_1^{[r,t]}=\sum_{l=0}^{r+t-2}
\sum_{m+n=t-1}^{\mbox{\tiny$\begin{array}{c}
0\leq m\leq l,\\
0\leq n\leq r+t-l-2\\
\end{array}$}}
\left(
\begin{array}{c}
l \\
m \\
\end{array}
\right)\dfrac{\ell^{*(l-m)}(-\ell)^{m}}{(\lambda_1-\lambda_1^{*})^{l+1}}
\Psi_1^{[r+t-l-n-2]\dag}\Psi_1^{[n]}.
$$

\subsection{Series expansions of triple-root case}

Assume $\rho=\upsilon=0$ in Eq. (\ref{10}) under a special spectral parameter $\lambda=\lambda_1$,
one can obtain a tripe root of the cubic spectral characteristic equation,
which can be explicitly given by $\xi_2=\xi_3=\xi_1=\frac{1}{6}\lambda_1+\frac{1}{3}\varsigma$.
In this circumstance,
we rewrite Eq. (\ref{lam}) as below to impose a small perturbation on the spectral parameter, viz.,
\begin{equation}\label{lam2}
\lambda_1(\epsilon)=\lambda_1+\ell\epsilon^3,
\end{equation}
where
$$
\ell=\dfrac{36}{4a_1^2-16 a_1 a_2+4 a_2^2-8\lambda_1^2-8\varsigma\lambda_1-3\varrho}.
$$
We thus get the expansion
\begin{equation}
\xi_{1}(\epsilon)=\sum_{j=0}^{\infty}\xi_1^{[j]}\epsilon^j,
\end{equation}
where
$$
\begin{array}{l}
\xi_{1}^{[0]}=\xi_1=\dfrac{1}{6}\lambda_1+\dfrac{1}{3}\varsigma,\ \xi_{1}^{[1]}=1,\ \xi_{1}^{[2]}=\dfrac{1}{9}\ell(2\lambda_{1}+\varsigma),\
\xi_{1}^{[3]}=\dfrac{1}{6}\ell,\\
\xi_{1}^{[4]}=-\dfrac{1}{3} (\xi_1^{[2]})^3+\dfrac{1}{3}(\ell-6 \xi_1^{[3]}) \xi_1^{[2]}
+\dfrac{1}{9}\ell(2\lambda_1+\varsigma)\xi_1^{[3]}
-\dfrac{1}{18}\ell^2\varsigma-\dfrac{1}{9} \ell^2 \lambda_1,\\
\xi_{1}^{[5]}=\dfrac{1}{6}(\ell-6\xi_1^{[3]}) (\xi_1^{[2]})^2-2 \xi_1^{[2]} \xi_1^{[4]}
-(\xi_1^{[3]})^2+\dfrac{1}{3}\ell \xi_1^{[3]}+\dfrac{1}{9}\ell(
2\ell \lambda_1+\varsigma) \xi_1^{[4]}+\dfrac{1}{12} \ell^2,\\
\xi_{1}^{[6]}=0,\
 \xi_{1}^{[7]}=-(\xi_1^{[2]})^2 \xi_1^{[5]}+\dfrac{1}{3}(\ell-6\xi_1^{[3]}) \xi_1^{[2]}\xi_1^{[4]}
-\dfrac{1}{3}(\xi_1^{[3]})^3+\dfrac{1}{6} \ell (\xi_1^{[3]})^2
\\ \hspace{2.5cm}+\dfrac{1}{12}(\ell^2-24\xi_1^{[5]}) \xi_1^{[3]}-\dfrac{1}{24} \ell^3+\dfrac{1}{3}\ell \xi_1^{[5]}-(\xi_1^{[4]})^2,
\end{array}$$
and $\xi_{1}^{[j]}$ ($j\geq 8$) yield
$$
\begin{array}{l}
\xi_1^{[j]}=-\dfrac{1}{3}\left[\displaystyle\sum_{m+n+k=j+2}^{0\leq m,n,k\leq j-1}\xi_1^{[m]}\xi_1^{[n]}\xi_1^{[k]}-\left(\dfrac{1}{2}\lambda_1+\varsigma\right)\sum_{m+n=j+2}^{0\leq m,n\leq j-1}\xi_1^{[m]}\xi_1^{[n]}-\dfrac{1}{2}\ell\sum_{m+n=j-1}\xi_1^{[m]}\xi_1^{[n]}\right.\\
~~~~~~~\left.-\dfrac{1}{2}\ell\lambda_1\xi_1^{[j-1]}-\dfrac{1}{4}\ell^2\xi_1^{[j-4]}\right].
\end{array}$$
In accordance with Eq. (\ref{Aj}), we can present the expansion
\begin{equation}\label{26}
A_1(\epsilon)=\sum_{j=0}^{\infty}A_1^{[j]}\epsilon^j,
\end{equation}
where
$$
A_1^{[j]}={\rm i}\left(\xi_1^{[j]}x+\dfrac{1}{\delta}\left[\sum_{k=0}^{[\frac{j}{3}]}\sum_{l=0}^{j-3k}
\dfrac{(-\ell)^k\sigma}{(\lambda_1-2\omega)^{k+1}}
\xi_1^{[l]}\xi_1^{[j-3k-l]}
+\sum_{k=0}^{[\frac{j}{3}]}\dfrac{(-\ell)^k(a_1b_2-a_2b_1)}{(\lambda_1-2\omega)^{k+1}}\xi_1^{[j-3k]} \right]t\right).
$$
Then by making use of the Schur polynomials one obtains
\begin{equation}\label{27}
\exp\left(\sum_{j=1}^{\infty}A_1^{[j]}\epsilon^j\right)
=\sum_{j=0}^{\infty}S^{[j]}(\mathbf{A})\epsilon^j,
\end{equation}
where $\mathbf{A}=(A_1^{[1]},A_1^{[2]},\cdots)$.

Moreover, it can be computed that
\begin{equation}\label{28}
\dfrac{1}{2\xi_{1}(\epsilon)-\lambda_1(\epsilon)-2a_{l}}=\sum_{j=0}^{\infty}
\mu_{l}^{[j]}\epsilon^j,\ l=1,2,
\end{equation}
where
$$
\begin{array}{l}
\mu_{l}^{[0]}=\dfrac{-1}{-\lambda_1-2a_{l}+2\xi_1},\ \mu_{l}^{[1]}=\dfrac{-2}{(-\lambda_1-2a_{l}+2\xi_1)^2},\
\mu_{l}^{[2]}=\dfrac{2\xi_1^{[2]}(-\lambda_1-2a_{l}+2\xi_1)+4}
{(-\lambda_1-2a_{l}+2\xi_1)^3},\\
\mu_{l}^{[k]}=\displaystyle\dfrac{-1}{-\lambda_1-2a_{l}
+2\xi_1}\left[\sum_{m=1}^{k}2\mu_{l}^{[k-m]}\xi_1^{[m]}
-\ell\mu_{l}^{[k-3]}  \right],\ k\geq 3.
\end{array}$$
Thus, in the light of Eqs. (\ref{27}) and (\ref{28}), we can introduce
\begin{equation}
\Upsilon(\epsilon)=\sum_{j=0}^{\infty}\Upsilon^{[j]}\epsilon^j,
\end{equation}
where
$$
\Upsilon^{[j]}=\left(
                         \begin{array}{c}
                           S^{[j]}(\mathbf{A}) \\
                           {\rm i}c_1\displaystyle\sum^{j}_{k=0}\mu_{1}^{[j-k]}S^{[k]}(\mathbf{A}) \\
                           {\rm i}c_2\displaystyle\sum^{j}_{k=0}\mu_{2}^{[j-k]}S^{[k]}(\mathbf{A}) \\
                         \end{array}
                       \right).
$$
Afterwards, we denote
\begin{equation}\label{30}
\begin{array}{l}
\Psi_1(\epsilon)=\Gamma_1
\left[\dfrac{\Upsilon(\epsilon)+\Upsilon(\varpi\epsilon)
+\Upsilon(\varpi^*\epsilon)}{3}\right]
+\Gamma_2\left[\dfrac{\Upsilon(\epsilon)
+\varpi^*\Upsilon(\varpi\epsilon)+\varpi\Upsilon(\varpi^*\epsilon)}{3\epsilon}\right]\\
\hspace{1.2cm}+\Gamma_3
\left[\dfrac{\Upsilon(\epsilon)+\varpi\Upsilon(\varpi\epsilon)
+\varpi^*\Upsilon(\varpi^*\epsilon)}{3\epsilon^2}\right],\end{array}
\end{equation}
where
$\Gamma_1=\sum_{k=0}^{\infty}\gamma_{3k}\epsilon^{3k},\ \Gamma_2=\sum_{k=0}^{\infty}\gamma_{3k+1}\epsilon^{3k},\
\Gamma_3=\sum_{k=0}^{\infty}\gamma_{3k+2}\epsilon^{3k}$, $\gamma_{j}$ $(j\geq 0)$ are complex constants and
$\varpi=\exp({\rm i}\frac{2}{3}\pi)$.
By direct calculation, Eq. (\ref{30}) can lead to the following expansion
associated with the terms in powers of $\epsilon^3$:
\begin{equation}\label{31}
\Psi_1(\epsilon)=\sum_{j=0}^{\infty}\Psi_1^{[j]}\epsilon^{3j},
\end{equation}
where
$$
\Psi_1^{[j]}=\sum_{k=0}^{j}\left[\Gamma_{3k}\Upsilon^{[3(j-k)]}
+\Gamma_{3k+1}
\Upsilon^{[3(j-k)+1]}+\Gamma_{3k+2}\Upsilon^{[3(j-k)+2]}\right].
$$
Here $\Psi_1^{[j]}=(\psi_1^{[j]},\varphi_1^{[j]},\chi_1^{[j]})^T$ and $G\Psi_1$ satisfies System (\ref{LP}) under
the plane-wave solution (\ref{08}) and the spectral parameter (\ref{lam2}).

Until now, we rewrite Eqs. (\ref{21}) and (\ref{22}) as
\begin{equation}
\dfrac{1}{\lambda_1(\epsilon)-\lambda_1(\epsilon)^{*}}
=\sum_{j=0}^{\infty}\sum_{m=0}^{j}\left(
\begin{array}{c}
j \\
m \\
\end{array}
\right)\dfrac{\ell^{*(j-m)}(-\ell)^{m}}
{(\lambda_1-\lambda_1^{*})^{j+1}}\epsilon^{*3(j-m)}\epsilon^{3m},
\end{equation}
and
\begin{equation}
\Psi_1(\epsilon)^{\dag}\Psi_1(\epsilon)
=\sum_{j=0}^{\infty}\sum_{m=0}^{j}\Psi_1^{[j-m]\dag}
\Psi_1^{[m]}\epsilon^{*3(j-m)}\epsilon^{3m}.
\end{equation}
Then it holds that
$$
\begin{array}{l}
\dfrac{\Psi_1(\epsilon)^{\dag}\Psi_1(\epsilon)}
{\lambda_1(\epsilon)-\lambda_1(\epsilon)^{*}}=
\displaystyle\sum_{j=0}^{\infty}\sum_{l=0}^{j}\left[\sum_{m=0}^{l}
\left(
\begin{array}{c}
l \\
m \\
\end{array}
\right)\dfrac{\ell^{*(l-m)}(-\ell)^{m}}
{(\lambda_1-\lambda_1^{*})^{l+1}}\epsilon^{*3(l-m)}\epsilon^{3m}\right]
\left[\sum_{n=0}^{j-l}\Psi_1^{[j-l-n]\dag}\Psi_1^{[n]}
\epsilon^{*3(j-l-n)}\epsilon^{3n}\right]\\
\hspace{2.7cm}=\displaystyle\sum_{j=0}^{\infty}\sum_{s=0}^{j}\left[\sum_{l=0}^{j}
\sum_{m+n=s}^{\mbox{\tiny$\begin{array}{c}
0\leq m\leq l,\\
0\leq n\leq j-l\\
\end{array}$}}
\left(
\begin{array}{c}
l \\
m \\
\end{array}
\right)\dfrac{\ell^{*(l-m)}(-\ell)^{m}}{(\lambda_1-\lambda_1^{*})^{l+1}}
\Psi_1^{[j-l-n]\dag}\Psi_1^{[n]}\right]\epsilon^{*3(j-s)}\epsilon^{3s}.
\end{array}$$
We thus end up with
\begin{equation}\label{34}
\dfrac{\Psi_1(\epsilon)^{\dag}\Psi_1(\epsilon)}
{\lambda_1(\epsilon)-\lambda_1(\epsilon)^{*}}=
\sum_{r,t=1}^{\infty}H_1^{[r,t]}\epsilon^{*3(r-1)}\epsilon^{3(t-1)},
\end{equation}
where
$$
H_1^{[r,t]}=\sum_{l=0}^{r+t-2}
\sum_{m+n=t-1}^{\mbox{\tiny$\begin{array}{c}
0\leq m\leq l,\\
0\leq n\leq r+t-l-2\\
\end{array}$}}
\left(
\begin{array}{c}
l \\
m \\
\end{array}
\right)\dfrac{\ell^{*(l-m)}(-\ell)^{m}}{(\lambda_1-\lambda_1^{*})^{l+1}}
\Psi_1^{[r+t-l-n-2]\dag}\Psi_1^{[n]}.
$$

\subsection{The $n$th-order rogue wave solution}

Given the series expansions of Eqs. (\ref{psi1}) and (\ref{23}) expressed in powers of $\epsilon^2$
or Eqs. (\ref{31}) and (\ref{34}) expressed in powers of $\epsilon^3$,
we can obatin the general $n$th-order rogue wave solution for the CMB equations in a unified
compact determinant form
\begin{subequations}\label{}
\begin{align}
&E_{1}[n]=\left(c_1+2{\rm i}\dfrac{1}{|H_1|}\left|
                                \begin{array}{cc}
                                  H_1 & Y_1^{\dag} \\
                                  X_1 & 0 \\
                                \end{array}
                              \right|\right){\rm e}^{{\rm i}\theta_1},\\
&E_{2}[n]=\left(c_2+2{\rm i}\dfrac{1}{|H_1|}\left|
                                \begin{array}{cc}
                                  H_1 & Z_1^{\dag}\\
                                  X_1 & 0 \\
                                \end{array}
                              \right|\right){\rm e}^{{\rm i}\theta_2},\\
&N[n]=N_0-2{\rm i}\dfrac{\partial}{\partial t}\dfrac{1}{|H_1|}\left|
                                \begin{array}{cc}
                                  H_1 & X_1^{\dag} \\
                                  X_1 & 0 \\
                                \end{array}
                              \right|,\\
&p_{1}[n]={\rm i}c_1b_1{\rm e}^{{\rm i}\theta_1}+2{\rm i}\dfrac{\partial}{\partial t}\left(\dfrac{1}{|H_1|}\left|
                                \begin{array}{cc}
                                  H_1 & Y_1^{\dag} \\
                                  X_1 & 0 \\
                                \end{array}
                              \right|{\rm e}^{{\rm i}\theta_1}\right),\\
&p_{2}[n]={\rm i}c_2b_2{\rm e}^{{\rm i}\theta_2}+2{\rm i}\dfrac{\partial}{\partial t}\left(\dfrac{1}{|H_1|}\left|
                                \begin{array}{cc}
                                  H_1 & Z_1^{\dag} \\
                                  X_1 & 0 \\
                                \end{array}
                              \right|{\rm e}^{{\rm i}\theta_2}\right),\\
&M_{11}[n]=m_{1}+2{\rm i}\dfrac{\partial}{\partial t}\left(\dfrac{1}{|H_1|}\left|
                                \begin{array}{cc}
                                  H_1 & Y_1^{\dag} \\
                                  Y_1 & 0 \\
                                \end{array}
                               \right|\right),\\
&M_{12}[n]=
-\dfrac{c_{1}c_{2}\sigma}{2\delta}{\rm e}^{{\rm i}(\theta_{2}-\theta_{1})}
+2{\rm i}\dfrac{\partial}{\partial t}\left(\dfrac{1}{|H_1|}\left|
                                \begin{array}{cc}
                                  H_1 & Z_1^{\dag} \\
                                  Y_1 & 0 \\
                                \end{array}
                              \right|{\rm e}^{{\rm i}(\theta_{2}-\theta_{1})}\right),\\
&M_{21}[n]=
-\dfrac{c_{1}c_{2}\sigma}{2\delta}{\rm e}^{{\rm i}(\theta_{1}-\theta_{2})}
+2{\rm i}\dfrac{\partial}{\partial t}\left(\dfrac{1}{|H_1|}\left|
                                \begin{array}{cc}
                                  H_1 & Y_1^{\dag} \\
                                  Z_1 & 0 \\
                                \end{array}
                              \right|{\rm e}^{{\rm i}(\theta_{1}-\theta_{2})}\right),\\
&M_{22}[n]=m_{2}+2{\rm i}\dfrac{\partial}{\partial t}\left(\dfrac{1}{|H_1|}\left|
                                \begin{array}{cc}
                                  H_1 & Z_1^{\dag} \\
                                  Z_1 & 0 \\
                                \end{array}
                              \right|\right),
\end{align}
\end{subequations}
where
$$\begin{array}{l}
H_{1}=\left(H_1^{[r,t]}\right)_{1\leq r,t\leq n},\\
X_1=(\psi_1^{[0]},\psi_1^{[1]},\cdots,\psi_1^{[n-1]}),\\
Y_1=(\varphi_1^{[0]},\varphi_1^{[1]},\cdots,\varphi_1^{[n-1]}),\\
Z_1=(\chi_1^{[0]},\chi_1^{[1]},\cdots,\chi_1^{[n-1]}).
\end{array}$$
Particularly, the first-order rogue wave solution can be calculated as
\begin{subequations}\label{r36}
\begin{align}
&E_{1}[1]=\left(c_1-2{\rm i}\dfrac{\psi_1^{[0]}\varphi_1^{[0]*}}{H_1^{[1,1]}}\right){\rm e}^{{\rm i}\theta_1},\\
&E_{2}[1]=\left(c_2-2{\rm i}\dfrac{\psi_1^{[0]}\chi_1^{[0]*}}{H_1^{[1,1]}}\right){\rm e}^{{\rm i}\theta_2},\\
&N[1]=N_0+2{\rm i}\dfrac{\psi_1^{[0]}\psi_1^{[0]*}}{H_1^{[1,1]}},\\
&p_{1}[1]={\rm i}c_1b_1{\rm e}^{i\theta_1}-2{\rm i}\dfrac{\partial}{\partial t}\left(\dfrac{\psi_1^{[0]}\varphi_1^{[0]*}}{H_1^{[1,1]}}{\rm e}^{{\rm i}\theta_1}\right),\\
&p_{2}[1]={\rm i}c_2b_2{\rm e}^{i\theta_2}-2{\rm i}\dfrac{\partial}{\partial t}\left(\dfrac{\psi_1^{[0]}\chi_1^{[0]*}}{H_1^{[1,1]}}{\rm e}^{{\rm i}\theta_2}\right),\label{p2}\\
&M_{11}[1]=m_{1}-2{\rm i}\dfrac{\partial}{\partial t}\left(\dfrac{\varphi_1^{[0]}\varphi_1^{[0]*}}{H_1^{[1,1]}}\right),\\
&M_{12}[1]=
-\dfrac{c_{1}c_{2}\sigma}{2\delta}{\rm e}^{{\rm i}(\theta_{2}-\theta_{1})}
-2{\rm i}\dfrac{\partial}{\partial t}\left(\dfrac{\varphi_1^{[0]}\chi_1^{[0]*}}{H_1^{[1,1]}}{\rm e}^{{\rm i}(\theta_{2}-\theta_{1})}\right),\label{M12}\\
&M_{21}[1]=
-\dfrac{c_{1}c_{2}\sigma}{2\delta}{\rm e}^{{\rm i}(\theta_{1}-\theta_{2})}
-2{\rm i}\dfrac{\partial}{\partial t}\left(\dfrac{\chi_1^{[0]}\varphi_1^{[0]*}}{H_1^{[1,1]}}{\rm e}^{{\rm i}(\theta_{1}-\theta_{2})}\right),\\
&M_{22}[1]=m_{2}-2{\rm i}\dfrac{\partial}{\partial t}\left(\dfrac{\chi_1^{[0]}\chi_1^{[0]*}}{H_1^{[1,1]}}\right),
\end{align}
\end{subequations}
and the second-order rogue wave solution is obtained as
\begin{subequations}\label{r37}
\begin{align}
&E_{1}[2]=\left[c_1+2{\rm i}\dfrac{\varphi_1^{[0]*}(\psi_1^{[1]}H_1^{[2,1]}-\psi_1^{[0]}H_1^{[2,2]})
+\varphi_1^{[1]*}(\psi_1^{[0]}H_1^{[1,2]}-\psi_1^{[1]}H_1^{[1,1]})}
{H_1^{[1,1]}H_1^{[2,2]}-H_1^{[1,2]}H_1^{[2,1]}}\right]{\rm e}^{{\rm i}\theta_1},\\
&E_{2}[2]=\left[c_2+2{\rm i}\dfrac{\chi_1^{[0]*}(\psi_1^{[1]}H_1^{[2,1]}-\psi_1^{[0]}H_1^{[2,2]})+
\chi_1^{[1]*}(\psi_1^{[0]}H_1^{[1,2]}-\psi_1^{[1]}H_1^{[1,1]})}
{H_1^{[1,1]}H_1^{[2,2]}-
H_1^{[1,2]}H_1^{[2,1]}}\right]{\rm e}^{{\rm i}\theta_2},\\
&N[2]=N_0-2{\rm i}\dfrac{
\psi_1^{[0]*}(\psi_1^{[1]}H_1^{[2,1]}-\psi_1^{[0]}H_1^{[2,2]})
+\psi_1^{[1]*}(\psi_1^{[0]}H_1^{[1,2]}-\psi_1^{[1]}H_1^{[1,1]})}{H_1^{[1,1]}H_1^{[2,2]}-
H_1^{[1,2]}H_1^{[2,1]}},\\
&p_{1}[2]={\rm i}c_1b_1{\rm e}^{{\rm i}\theta_1}+2{\rm i}\dfrac{\partial}{\partial t}\left[\dfrac
{\varphi_1^{[0]*}(\psi_1^{[1]}H_1^{[2,1]}-\psi_1^{[0]}H_1^{[2,2]})
+\varphi_1^{[1]*}(\psi_1^{[0]}H_1^{[1,2]}-\psi_1^{[1]}H_1^{[1,1]})}
{H_1^{[1,1]}H_1^{[2,2]}-
H_1^{[1,2]}H_1^{[2,1]}}{\rm e}^{{\rm i}\theta_1}\right],\\
&p_{2}[2]={\rm i}c_2b_2{\rm e}^{{\rm i}\theta_2}+2{\rm i}\dfrac{\partial}{\partial t}\left[\dfrac{
\chi_1^{[0]*}(\psi_1^{[1]}H_1^{[2,1]}-\psi_1^{[0]}H_1^{[2,2]})
+\chi_1^{[1]*}(\psi_1^{[0]}H_1^{[1,2]}-\psi_1^{[1]}H_1^{[1,1]})
}{H_1^{[1,1]}H_1^{[2,2]}-
H_1^{[1,2]}H_1^{[2,1]}}{\rm e}^{{\rm i}\theta_2}\right],\\
&M_{11}[2]=m_{1}+2{\rm i}\dfrac{\partial}{\partial t}\left[\dfrac
{\varphi_1^{[0]*}(\varphi_1^{[1]}H_1^{[2,1]}-\varphi_1^{[0]}H_1^{[2,2]})
+\varphi_1^{[1]*}(\varphi_1^{[0]}H_1^{[1,2]}-\varphi_1^{[1]}H_1^{[1,1]})}{H_1^{[1,1]}H_1^{[2,2]}-
H_1^{[1,2]}H_1^{[2,1]}}\right],\\
&M_{12}[2]=
-\dfrac{c_{1}c_{2}\sigma}{2\delta}{\rm e}^{{\rm i}(\theta_{2}-\theta_{1})}
+2{\rm i}\dfrac{\partial}{\partial t}\left[\dfrac{\chi_1^{[0]*}(\varphi_1^{[1]}H_1^{[2,1]}
-\varphi_1^{[0]}H_1^{[2,2]})
+\chi_1^{[1]*}(\varphi_1^{[0]}H_1^{[1,2]}
-\varphi_1^{[1]}H_1^{[1,1]})}{H_1^{[1,1]}H_1^{[2,2]}-
H_1^{[1,2]}H_1^{[2,1]}}{\rm e}^{{\rm i}(\theta_{2}-\theta_{1})}\right],\\
&M_{21}[2]=-\dfrac{c_{1}c_{2}\sigma}{2\sigma}{\rm e}^{{\rm i}(\theta_{1}-\theta_{2})}
+2{\rm i}\dfrac{\partial}{\partial t}\left[\dfrac{\varphi_1^{[0]*}(\chi_1^{[1]}H_1^{[2,1]}
-\chi_1^{[0]}H_1^{[2,2]})
+\varphi_1^{[1]*}(\chi_1^{[0]}H_1^{[1,2]}
-\chi_1^{[1]}H_1^{[1,1]})}{H_1^{[1,1]}H_1^{[2,2]}-
H_1^{[1,2]}H_1^{[2,1]}}{\rm e}^{{\rm i}(\theta_{1}-\theta_{2})}\right],\\
&M_{22}[2]=m_{2}+2{\rm i}\dfrac{\partial}{\partial t}\left[\dfrac{
\chi_1^{[0]*}(\chi_1^{[1]}H_1^{[2,1]}-\chi_1^{[0]}H_1^{[2,2]})
+\chi_1^{[1]*}(\chi_1^{[0]}H_1^{[1,2]}-\chi_1^{[1]}H_1^{[1,1]})}{H_1^{[1,1]}H_1^{[2,2]}-
H_1^{[1,2]}H_1^{[2,1]}}\right].
\end{align}
\end{subequations}
where
$$
\begin{array}{l}
H_1^{[1,1]}=\dfrac{\Psi_1^{[0]\dag}\Psi_1^{[0]}}{\lambda_1-\lambda_1^{*}},
\ H_1^{[1,2]}=\dfrac{1}{\lambda_1-\lambda_1^{*}}\left[\Psi_1^{[0]\dag}\Psi_1^{[1]}-\dfrac{\Psi_1^{[0]\dag}\Psi_1^{[0]}
\ell}{\lambda_1-\lambda_1^{*}}\right],\\
H_1^{[2,1]}=\dfrac{1}{\lambda_1-\lambda_1^{*}}\left[\Psi_1^{[1]\dag}\Psi_1^{[0]}+\dfrac{\Psi_1^{[0]\dag}\Psi_1^{[0]}
\ell^{*}}{\lambda_1-\lambda_1^{*}}\right],\\
H_1^{[2,2]}=\dfrac{1}{\lambda_1-\lambda_{1}^{*}}
\left[\Psi_1^{[1]\dag}\Psi_1^{[1]}-\dfrac{\left((\lambda_1-\lambda_1^{*})\Psi_1^{[1]\dag}\Psi_1^{[0]}+\Psi_1^{[0]\dag}
\Psi_1^{[0]}\ell^{*}\right)\ell}
{(\lambda_1-\lambda_{1}^{*})^2}\right.\\
~~~~\left.+\dfrac{\left((\lambda_1-\lambda_1^{*})\Psi_1^{[0]\dag}\Psi_1^{[1]}-\Psi_1^{[0]\dag}\Psi_1^{[0]}\ell\right)\ell^{*}}
{(\lambda_1-\lambda_{1}^{*})^2}\right].
\end{array}$$

It is well known that the coupled systems could be provided with
more diverse and complex rogue wave structures than the scalar ones.
Therefore, including nine components in Eqs. (\ref{e21})-(\ref{e29}),
the CMB equations might admit some extraordinary rogue wave structures apart from the
Peregrine soliton, dark and four-petaled rogue waves that can be attainable
in the coupled NLS equations. Next we will show some unusual rogue wave dynamics
such as triple-hole and twisted-pair rogue waves as well as some new composite rogue waves
through Eqs. (\ref{r36}) and (\ref{r37}) by discussing the aforementioned two cases.

Case 1. The double-root case. In this situation, we give a concrete example to
reveal the rogue wave dynamics. For instance, we choose
\begin{align}
&c_1=c_2=1,a_1=-a_2=1,m_1=m_2=1,\ N_0=1,\nonumber\\
&\gamma_0=0,\gamma_1=1,b_1=-\dfrac{4}{13},b_2=-\dfrac{12}{13},\omega=1,\nonumber\\
&\lambda_1=\dfrac{1}{8}\sqrt{24\sqrt{6}-9}-{\rm i}\dfrac{1}{8}\sqrt{24\sqrt{6}+9},\nonumber\\
&\xi_2=\xi_1=\left(-\dfrac{1}{20}-\dfrac{3}{160}\sqrt{6}-{\rm i}\dfrac{11}{800}\sqrt{10}
+{\rm i}\dfrac{\sqrt{15}}{100} \right)\sqrt{16\sqrt{6}-6},\nonumber
\end{align}
then through Eqs. (\ref{psi1}), (\ref{23}) and (\ref{r36}),
we can obtain different types of rogue wave structures in each component of the CMB equations,
see the following table:
\begin{center}
Table 1: types of rogue waves in each component for the double-root case.
\begin{tabular}{lcccccccc}
\hline
component &$E_1$ &$E_2$ &$N$ &$p_1$ &$p_2$ &$M_{11}$ &$M_{12}$/$M_{21}$ &$M_{22}$ \\
\hline
pattern   & dark  & bright & dark & four-petaled
& triple-hole & four-petaled
& twisted-pair & dark \\
\hline
\end{tabular}
\end{center}

\begin{figure}[!h]
\centering
\renewcommand{\figurename}{{\bf Fig.}}
{\includegraphics[height=4cm,width=6cm]{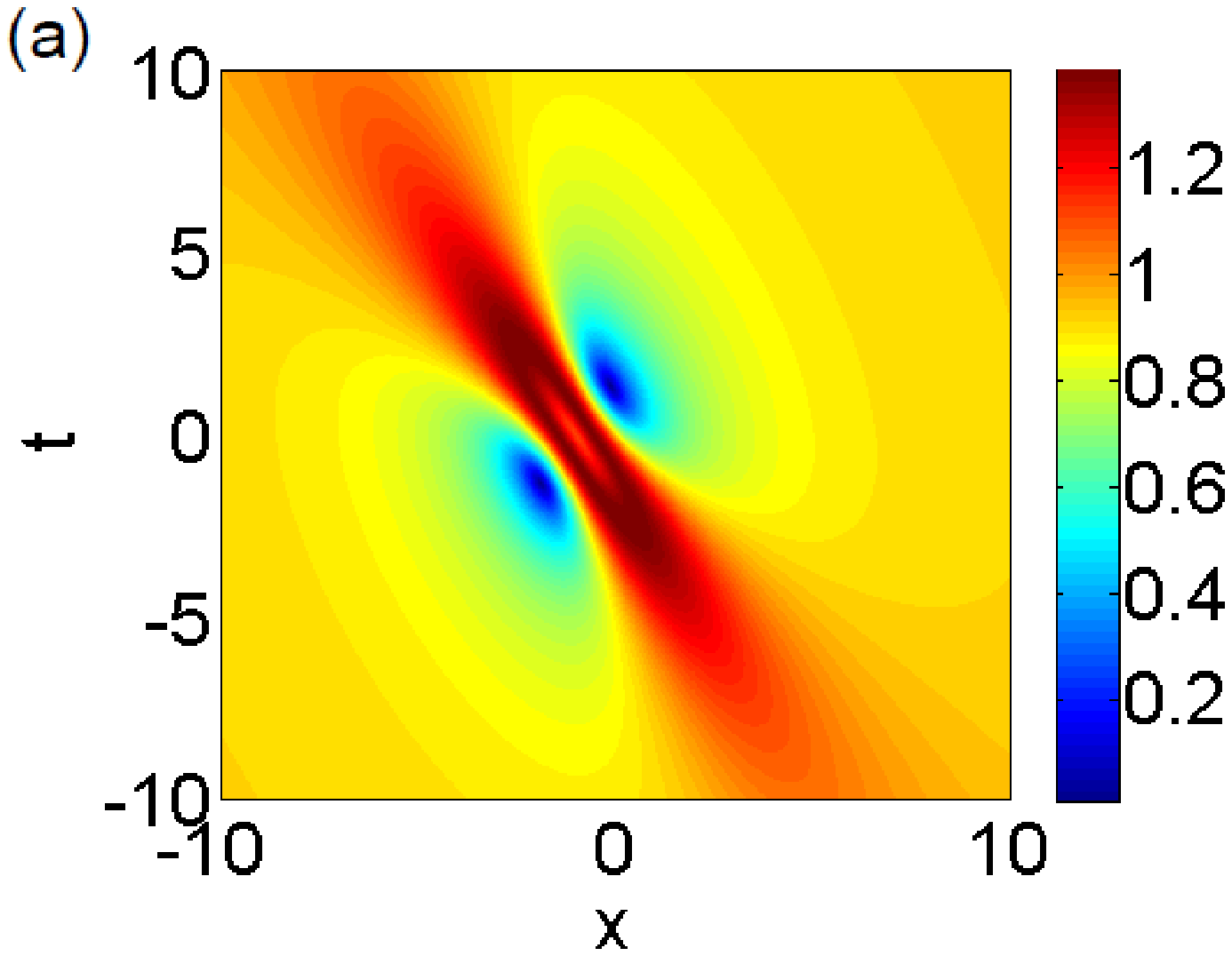}}
{\includegraphics[height=4cm,width=6cm]{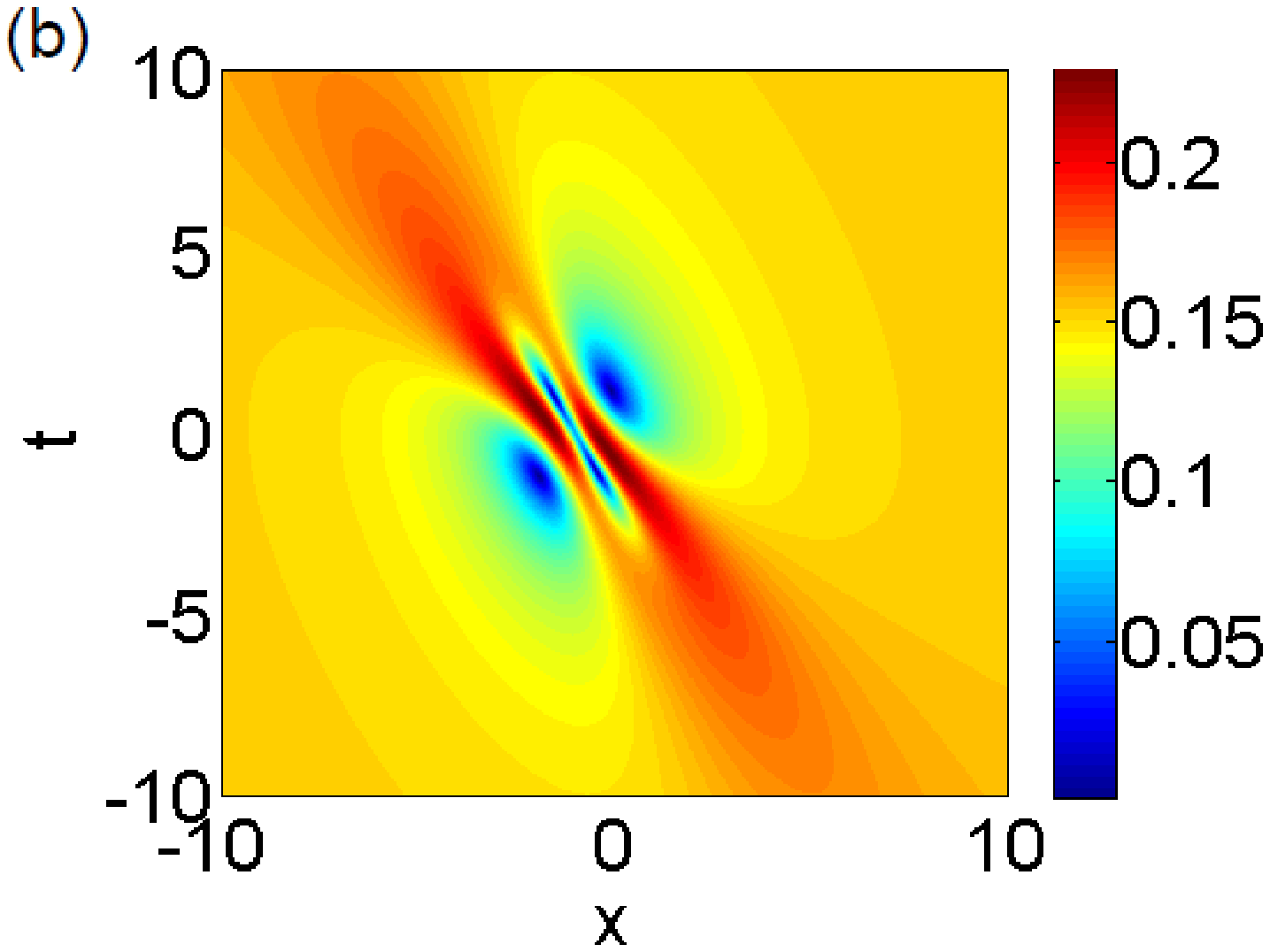}}
\caption{(a), (b) Density plots of the triple-hole rogue wave $|p_2[1]|$ and the twisted-pair rogue wave
$|M_{12}[1]|$, respectively.}
\label{fig:1}
\end{figure}

\begin{figure}[!h]
\centering
\renewcommand{\figurename}{{\bf Fig.}}
{\includegraphics[height=4cm,width=6cm]{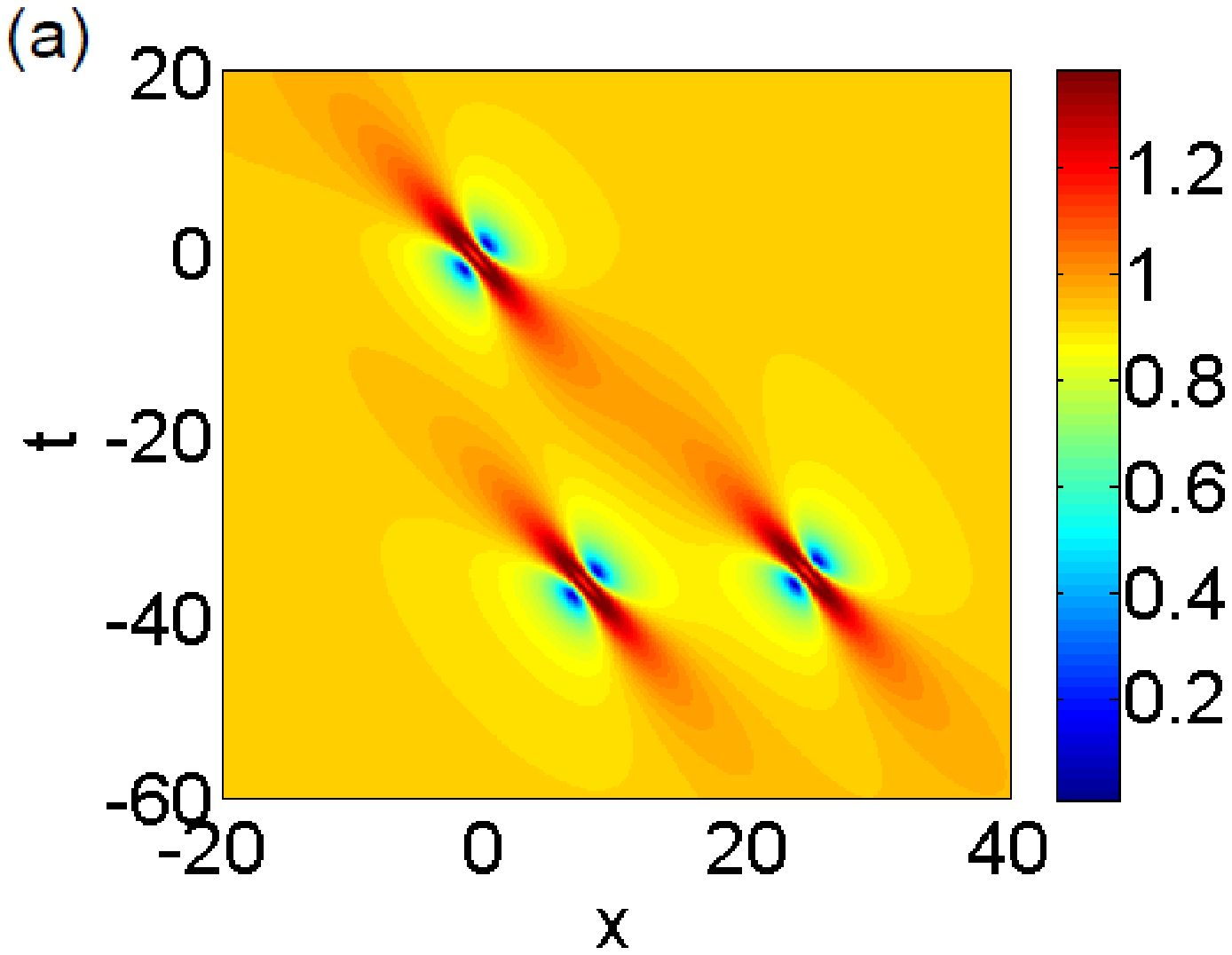}}
{\includegraphics[height=4cm,width=6cm]{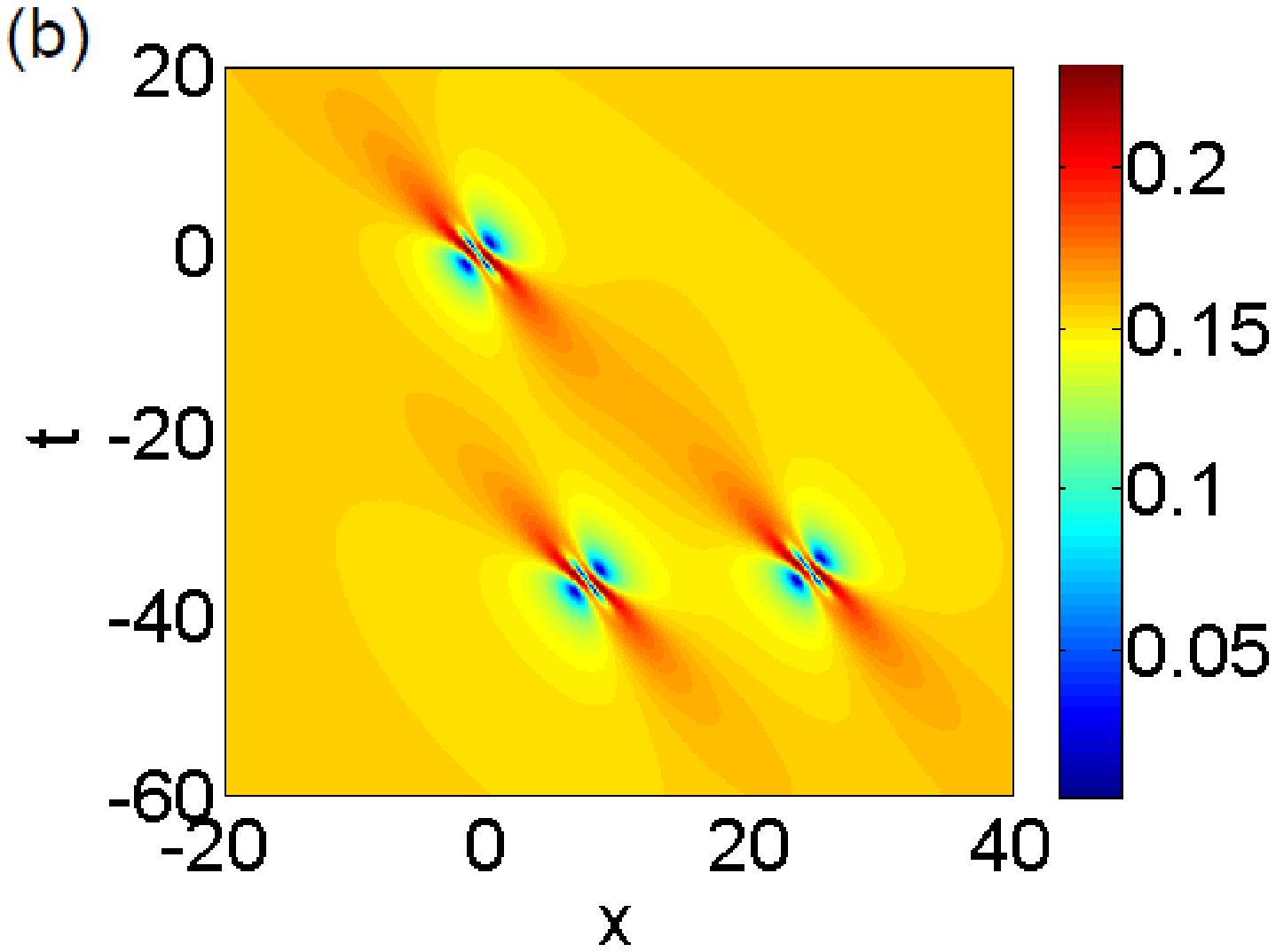}}
\caption{(a), (b) Triplet patterns of the triple-hole rogue wave
$|p_2[2]|$ and the twisted-pair rogue wave $|M_{12}[2]|$, respectively.}
\label{fig:2}
\end{figure}

As shown in Fig. \ref{fig:1}(a) and Fig. \ref{fig:1}(b)
the triple-hole rogue wave in the $p_2$ component and
the twisted-pair rogue wave in the $M_{12}$ component, respectively.
The three holes in the triple-hole rogue wave are localized in (-1.91,-1.33), (-0.14,1.33)
and (-1.02,-0.01). The twisted-pair rogue wave consists of two peaks which appear at
(-1.87,0.48) and (-0.19,-0.47) and four holes whose coordinates are (-1.94,-1.18),
(-1.56,1.02), (-0.49,-1.02) and (-0.11,1.18), respectively.
We should emphasize that, the twisted-pair rogue wave was first reported in
the Sasa--Satsuma equation \cite{42}, while the triple-hole rogue wave is one novel type of
rogue wave structure found in the CMB equations. To proceed, when choosing the parameters
$\gamma_0=10,\gamma_1=1,\gamma_2=0,\gamma_3=0$, the triplet rogue wave patterns
can be exhibited with the help of Eq. (\ref{r37}), see Fig. \ref{fig:2}.

Case 2. The triple-root case. In this circumstance, we set the parameters be
\begin{align}
&c_1=c_2=1,a_1=-a_2=\dfrac{1}{4},m_1=m_2=1,\ N_0=1,\nonumber\\
&\gamma_0=0,\gamma_1=1,\gamma_2=0,b_1=-\dfrac{28}{67},b_2=-\dfrac{36}{67},\omega=1,\nonumber\\
&\lambda_1={\rm i}\dfrac{3}{4}\sqrt{3},\xi_2=\xi_3=\xi_1={\rm i}\dfrac{1}{8}\sqrt{3}.\nonumber
\end{align}
Insertion of (\ref{31}) and (\ref{34}) into Eq. (\ref{r36}) under these
specific choices of the parameters we can arrive at the composite patterns of rogue waves in
each component of the CMB equations, which we list in the following table:
\begin{center}
Table 2: types of rogue waves in each component for the triple-root case.
\begin{tabular}{lcccccccc}
\hline
component &$E_1$ &$E_2$ &$N$ &$p_1$ &$p_2$ &$M_{11}$ &$M_{12}$/$M_{21}$ &$M_{22}$ \\
\hline
composite pattern
& bright  & bright & dark & bright
& four-petaled
& dark &bright
& four-petaled \\
\hline
\end{tabular}
\end{center}

\begin{figure}[!h]
\centering
\renewcommand{\figurename}{{\bf Fig.}}
{\includegraphics[height=4cm,width=6cm]{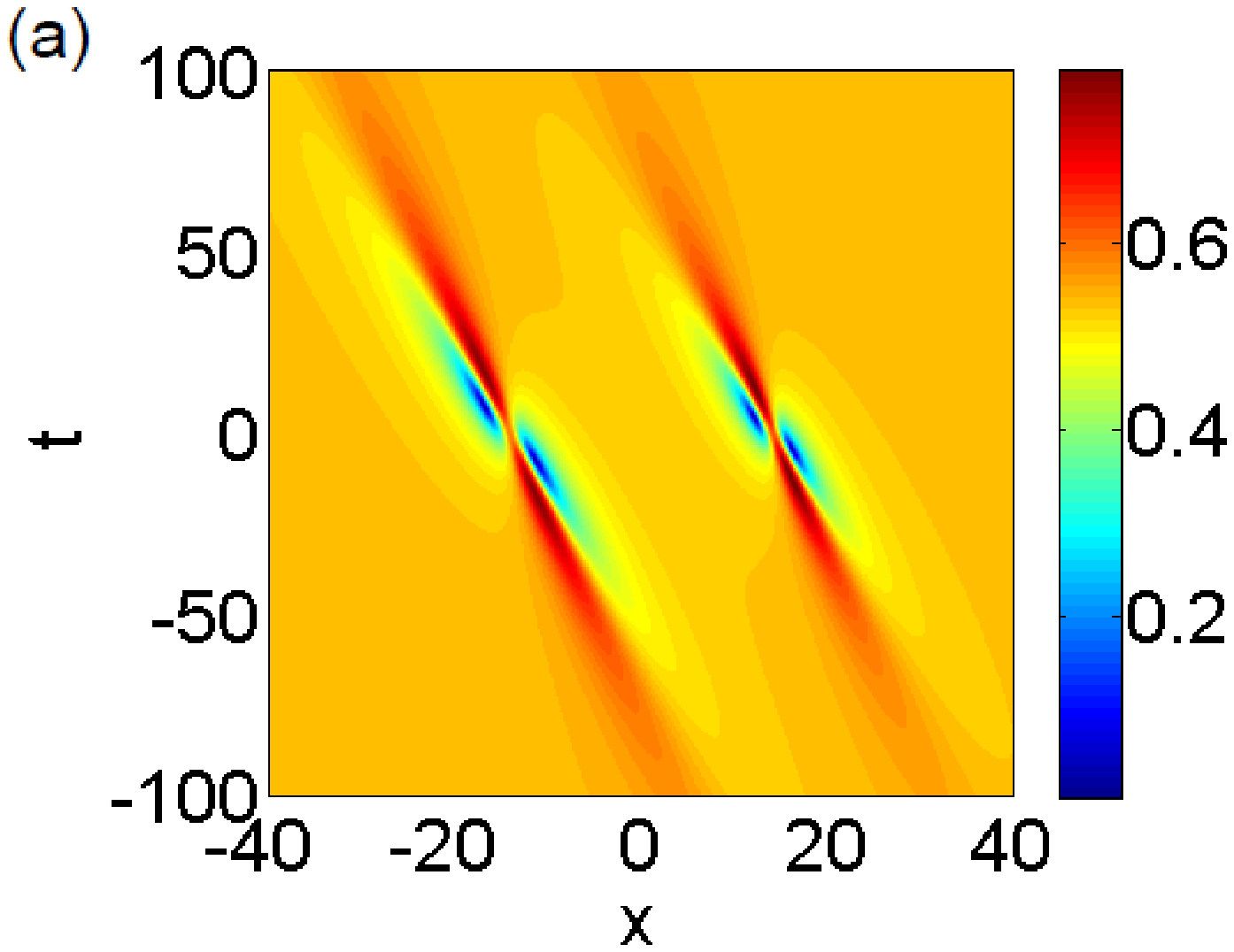}}
{\includegraphics[height=4cm,width=6cm]{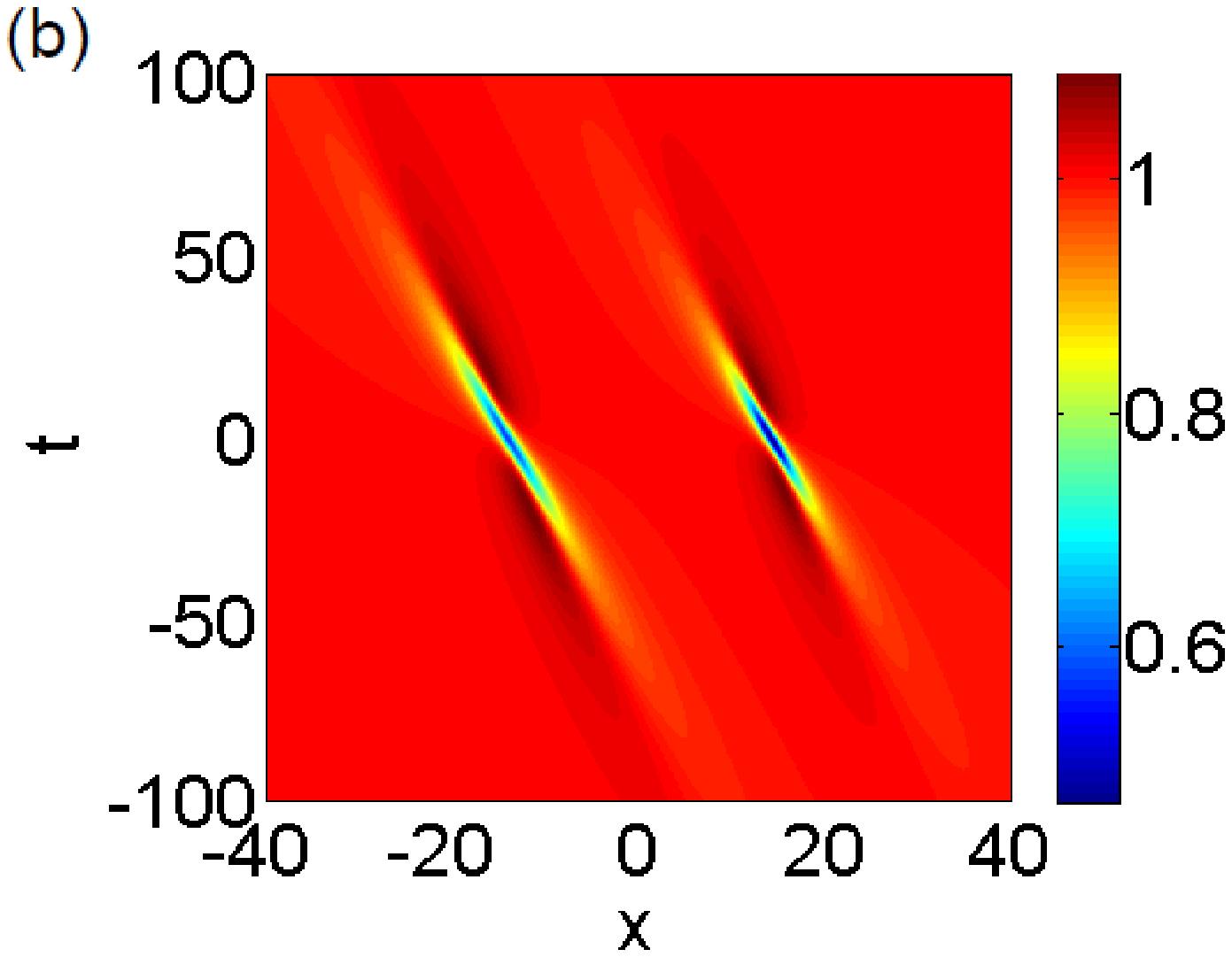}}
\caption{(a), (b) Density plots of the composite four-petaled rogue waves $|p_2[1]|$
and the composite dark rogue waves $|M_{11}[1]|$, respectively.}
\label{fig:3}
\end{figure}

\begin{figure}[!h]
\centering
\renewcommand{\figurename}{{\bf Fig.}}
{\includegraphics[height=4cm,width=6cm]{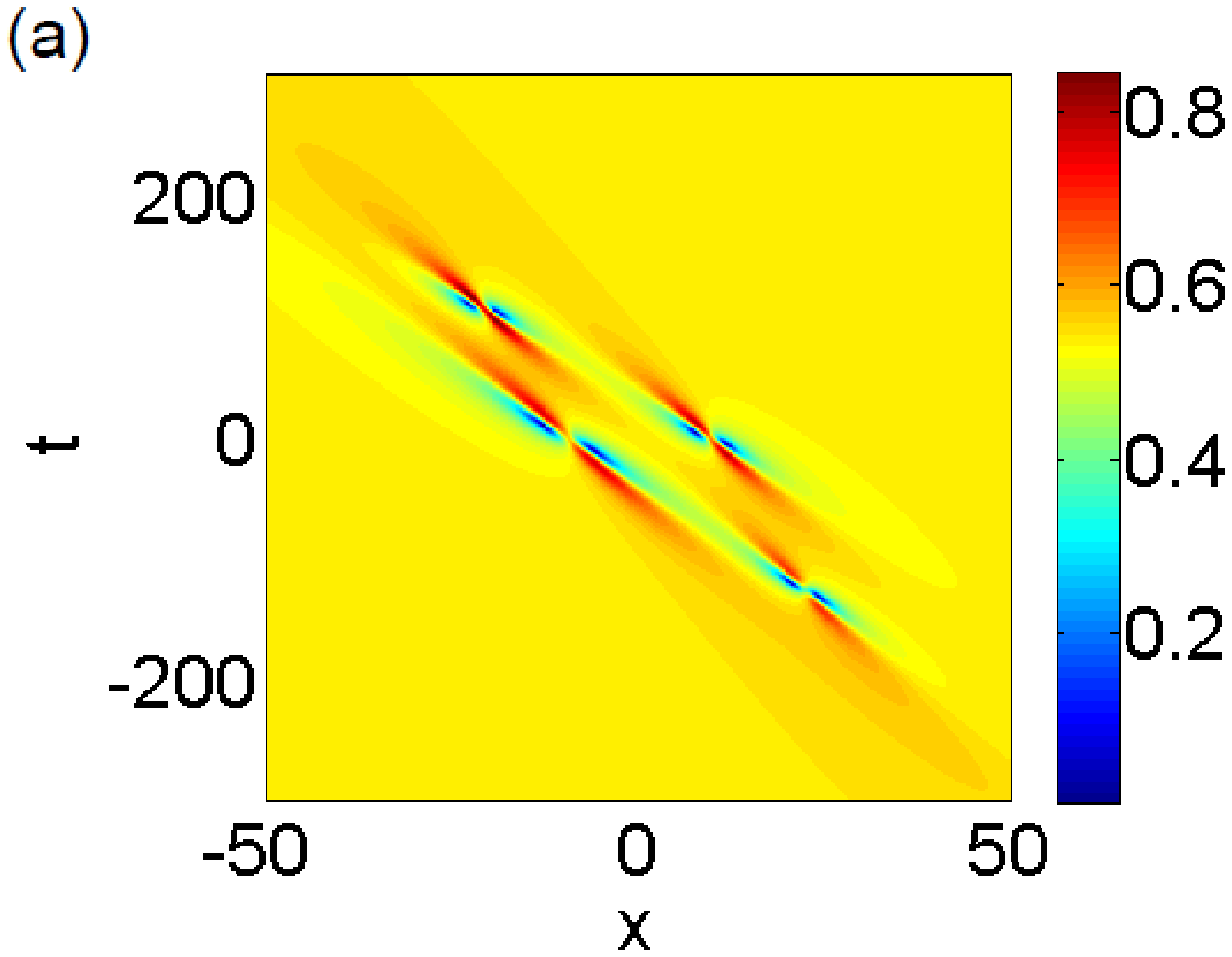}}
{\includegraphics[height=4cm,width=6cm]{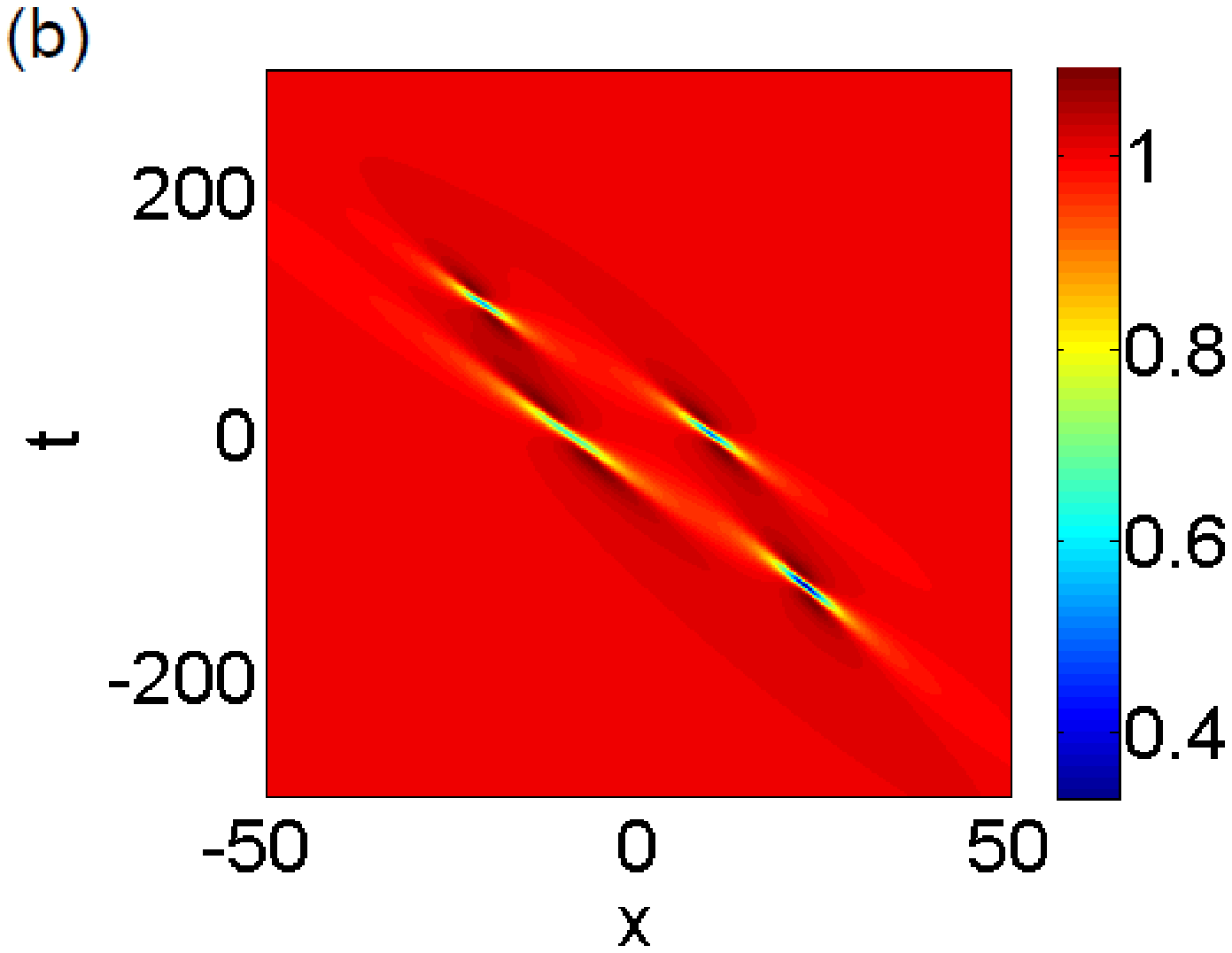}}
\caption{(a), (b) Quadruple patterns of the composite four-petaled rogue waves
$|p_2[2]|$ and the composite dark rogue waves $|M_{11}[2]|$, respectively.}
\label{fig:4}
\end{figure}

Fig. \ref{fig:3}(a) displays the composite four-petaled rogue waves in the $|p_2[1]|$ component.
The critical points are localized at (-16.88,7.63), (-16.90,17.09), (-11.31,-7.64), (-11.29,-17.41),
(12.07,4.35), (12.15,12.33), (16.12,-4.35) and (16.04,-12.32). Fig. \ref{fig:3}(b) illustrates the
composite dark rogue waves in the $M_{11}$ component. The two holes occur
at (-14.10,0.02) and (14.10,-0.02). Meanwhile, by taking advantage of Eq. (\ref{r37}) and choosing
suitable parameters, one can obtain the quadruple and sextuple composite rogue waves.
We show the quadruple pattern as an example by taking
$\gamma_0=100,\gamma_1=0,\gamma_2=1,\gamma_3=0,\gamma_4=0,\gamma_5=0$, see Fig. \ref{fig:4}.

\section{Semirational solutions}

To derive the semirational solutions, we let $a_2=a_1$ and $b_2=b_1$ in Eq. (\ref{08}),
then following plane-wave solution can be given, as
\begin{equation}\label{38}
\begin{array}{l}
\widehat{E}_{j}[0]=c_j{\rm e}^{{\rm i}\theta_{1}},\
\widehat{p}_{j}[0]={\rm i}c_jb_1{\rm e}^{{\rm i}\theta_{1}},\
\widehat{M}_{12}[0]=\widehat{M}_{21}[0]=0,\\
\widehat{N}[0]=N_0,\ \widehat{M}_{jj}[0]=-2a_1b_1-4b_1\omega-N_0,\ j=1,2.

\end{array}
\end{equation}
At this moment, the fundamental solution of Eqs. (\ref{IST}) and (\ref{IST2})
under this plane-wave solution reads
\begin{equation}
\widehat{\Phi}=\widehat{G}\left(
       \begin{array}{ccc}
         1 & 1 & 0 \\
         \dfrac{{\rm i}c_{1}}{2\widehat{\xi}_{1}-\lambda-2a_{1}} & \dfrac{{\rm i}c_{1}}{2\widehat{\xi}_{2}-\lambda-2a_{1}} & \alpha c_2 \\
         \dfrac{{\rm i}c_{2}}{2\widehat{\xi}_{1}-\lambda-2a_{1}} & \dfrac{{\rm i}c_{1}}{2\widehat{\xi}_{2}-\lambda-2a_{1}} & -\alpha c_1 \\
       \end{array}
     \right)\left(
              \begin{array}{c}
                {\rm e}^{\widehat{A}_{1}} \\
                {\rm e}^{\widehat{A}_{2}} \\
                {\rm e}^{B} \\
              \end{array}
            \right),
\end{equation}
where $\widehat{G}={\rm diag}(1,{\rm e}^{-{\rm i}\theta_{1}},{\rm e}^{-{\rm i}\theta_{1}})$, $\alpha$ is an arbitrary complex constant,
and $\widehat{\xi}_{j}$ ($j=1,2$) satisfy the following quadratic equation
\begin{equation}\label{40}
4\widehat{\xi}^2-4a_1\widehat{\xi}-\lambda^2-2a_1\lambda-\varrho=0,
\end{equation}
namely,
\begin{align}
&\widehat{\xi}_{1}=\dfrac{1}{2}a_1+\dfrac{1}{2}\sqrt{\lambda^2+2a_1\lambda+a_1^2+\varrho},\nonumber\\
&\widehat{\xi}_{2}=\dfrac{1}{2}a_1-\dfrac{1}{2}\sqrt{\lambda^2+2a_1\lambda+a_1^2+\varrho},\nonumber
\end{align}
and
\begin{align}
&\widehat{A}_j={\rm i}\left(\widehat{\xi}_j x+\dfrac{b_1}{\lambda-2\omega}\widehat{\xi}_j t\right),\ j=1,2,\\
&B={\rm i}\left[\left(\dfrac{\lambda}{2}+a_1\right)x+\dfrac{b_1}{\lambda-2\omega}\left(\dfrac{\lambda}{2}+a_1\right)t \right].
\end{align}

\subsection{Series expansions}

It is easy to see that Eq. (\ref{40}) has a double root $\xi_2=\xi_1=\frac{1}{2}a_1$
when $\lambda=\widehat{\lambda}_1=-a_1+{\rm i}\sqrt{\varrho}$. Hence suppose
\begin{equation}\label{43}
\widehat{\lambda}_1(\epsilon)=-a_1+{\rm i}\sqrt{\varrho}-\dfrac{2{\rm i}}{\sqrt{\varrho}}\epsilon^2,
\end{equation}
one obtains the expansion
\begin{equation}\label{44}
\widehat{\xi}_{1}(\epsilon)=\dfrac{1}{2}a_1+\sum_{j=1}^{\infty}\widehat{\xi}_1^{[j]}\epsilon^{2j-1},
\end{equation}
where
$$
\begin{array}{l}
\widehat{\xi}_1^{[1]}=1,\ \widehat{\xi}_1^{[2]}=-\dfrac{1}{2\varrho},\
\widehat{\xi}_1^{[j]}=-\displaystyle\dfrac{1}{2}\sum_{m+n=j+1}^{0\leq m,n\leq j-1}\widehat{\xi}_1^{[m]}\widehat{\xi}_1^{[n]},\ j\geq 3.
\end{array}$$
Using Eq. (\ref{43}) and (\ref{44}), we have
\begin{equation}\label{45}
\widehat{A}_1(\epsilon)=\widehat{A}_1^{[0]}+\sum_{j=1}^{\infty}\widehat{A}_1^{[j]}\epsilon^{2j-1},
\end{equation}
where
$$
\begin{array}{l}
\widehat{A}_1^{[0]}={\rm i}\left[\dfrac{1}{2}a_1 x+\dfrac{a_1b_1}{2(i\sqrt{\varrho}-a_1-2\omega)}t \right],\\
\widehat{A}_1^{[j]}=\displaystyle {\rm i}\left(\widehat{\xi}_1^{[j]}x
+b_1\left[\sum_{k=0}^{[\frac{2j-1}{2}]}\dfrac{(2i)^{k}}
{(i\sqrt{\varrho}-a_1-2\omega)^{k+1}(\sqrt{\varrho})^{k}}\widehat{\xi}_1^{[j-k]}\right]t\right),\ j\geq 1,
\end{array}$$
and
\begin{equation}\label{46}
B(\epsilon)=\sum_{j=0}^{\infty}B^{[j]}\epsilon^{2j},
\end{equation}
where
$$\begin{array}{l}
B^{[0]}={\rm i}\left[\left(\dfrac{a_1}{2}+\dfrac{{\rm i}\sqrt{\varrho}}{2}\right)x
+\dfrac{b_1}{({\rm i}\sqrt{\varrho}-a_1-2\omega)}\left(\dfrac{a_1}{2}
+\dfrac{{\rm i}\sqrt{\varrho}}{2}\right)t\right],\\
B^{[1]}={\rm i}\left(-\dfrac{{\rm i}}{\sqrt{\varrho}}x+\left[\dfrac{2{\rm i}b_1}{({\rm i}\sqrt{\varrho}-a_1-2\omega)^2\sqrt{\varrho}}
\left(\dfrac{a_1}{2}+\dfrac{{\rm i}\sqrt{\varrho}}{2}\right)
-\dfrac{b_1}{({\rm i}\sqrt{\varrho}-a_1-2\omega)}\dfrac{{\rm i}}{\sqrt{\varrho}}\right]t\right),\\
B^{[j]}={\rm i}\left[\dfrac{(2{\rm i})^{j}b_1}{({\rm i}\sqrt{\varrho}-a_1-2\omega)^{j+1}(\sqrt{\varrho})^{j}}
\left(\dfrac{a_1}{2}+\dfrac{{\rm i}\sqrt{\varrho}}{2}\right)
-\dfrac{(2{\rm i})^{j-1}b_1}{({\rm i}\sqrt{\varrho}-a_1-2\omega)^{j}}\dfrac{{\rm i}}{(\sqrt{\varrho})^{j}}\right]t,j\geq 2.
\end{array}$$
Next, by means of Eqs. (\ref{45}), (\ref{46}) and the Schur polynomials, we have
\begin{equation}
\exp\left(\widehat{A}_1^{[0]}+\sum_{j=1}^{\infty}\widehat{A}_1^{[j]}\epsilon^{2j-1}\right)
=\exp(\widehat{A}_1^{[0]})\left(\sum_{j=0}^{\infty}\widehat{S}^{[j]}(\mathbf{\widehat{A}})\epsilon^j\right),
\end{equation}
where
$\mathbf{\widehat{A}}=(\widehat{A}_1^{[1]},0,\widehat{A}_1^{[2]},0,\cdots),$
and
\begin{equation}
\exp\left(\sum_{j=0}^{\infty}B^{[j]}\epsilon^{2j}\right)
=\exp(B^{[0]})\left(\sum_{j=0}^{\infty}S^{[j]}(\mathbf{B})\epsilon^{2j}\right),
\end{equation}
where $\mathbf{B}=(B^{[1]},B^{[2]},\cdots).$

Moreover, considering
\begin{equation}
\dfrac{1}{2\widehat{\xi}_{1}(\epsilon)-
\widehat{\lambda}_1(\epsilon)-2a_{1}}=\sum_{j=0}^{\infty}\widehat{\mu}_{1}^{[j]}\epsilon^j,
\end{equation}
where
$$
\begin{array}{l}
\widehat{\mu}_{1}^{[0]}=-\dfrac{1}{{\rm i}\sqrt{\varrho}},\ \widehat{\mu}_{1}^{[1]}=\dfrac{2}{\varrho},\
\widehat{\mu}_{1}^{[k]}=\dfrac{1}{{\rm i}\sqrt{\varrho}}\displaystyle
\left[\sum_{m=0}^{[\frac{k+1}{2}]}2\widehat{\mu}_{1}^{[k+1-2m]}\widehat{\xi}_{1}^{[m]}
+\dfrac{2{\rm i}}{\sqrt{\varrho}}\widehat{\mu}_{1}^{[k-2]}\right],\ k\geq 2,
\end{array}$$
we can define
\begin{equation}
\widehat{\Upsilon}(\epsilon)=\sum_{j=0}^{\infty}\widehat{\Upsilon}^{[j]}\epsilon^j,
\end{equation}
where
$$
\widehat{\Upsilon}^{[j]}=\exp(\widehat{A}_1^{[0]})\left(
                         \begin{array}{c}
\widehat{S}^{[j]}(\mathbf{\widehat{A}}) \\
{\rm i}c_1\displaystyle\sum^{j}_{k=0}\widehat{\mu}_{1}^{[j-k]}\widehat{S}^{[k]}(\mathbf{\widehat{A}}) \\
{\rm i}c_2\displaystyle\sum^{j}_{k=0}\widehat{\mu}_{1}^{[j-k]}\widehat{S}^{[k]}(\mathbf{\widehat{A}}) \\
\end{array}
\right),
$$
and
\begin{equation}
\Theta(\epsilon)=\sum_{j=0}^{\infty}\Theta^{[j]}\epsilon^{2j},
\end{equation}
where
$$
\Theta^{[j]}=\exp(B^{[0]})\left(
               \begin{array}{c}
                 0 \\
                 \alpha c_2S^{[j]}(\mathbf{B}) \\
                 -\alpha c_1S^{[j]}(\mathbf{B}) \\
               \end{array}
             \right).
$$
In analogy ro Eqs. (\ref{19}) and (\ref{30}), we construct
\begin{equation}
\widehat{\Psi}_1(\epsilon)=\widehat{\Gamma}_1
\left[\dfrac{\widehat{\Upsilon}(\epsilon)+\widehat{\Upsilon}(-\epsilon)}{2}+\Theta(\epsilon)\right]
+\widehat{\Gamma}_2\left[\dfrac{\widehat{\Upsilon}(\epsilon)-\widehat{\Upsilon}
(-\epsilon)}{2\epsilon}+\Theta(\epsilon)\right],
\end{equation}
where
$\widehat{\Gamma}_1=\sum_{k=0}^{\infty}\widehat{\gamma}_{2k}\epsilon^{2k},\ \widehat{\Gamma}_2=\sum_{k=0}^{\infty}\widehat{\gamma}_{2k+1}\epsilon^{2k}$, $\widehat{\gamma}_{j}$
($j\geq 0$) are complex constants. Then $\widehat{\Psi}_1$ can be expanded as the terms
expressed in powers of $\epsilon^2$, that is
\begin{equation}\label{53}
\widehat{\Psi}_1(\epsilon)=\sum_{j=0}^{\infty}\widehat{\Psi}_1^{[j]}\epsilon^{2j},
\end{equation}
where
$$
\widehat{\Psi}_1^{[j]}=\sum_{k=0}^{j}\left[\widehat{\gamma}_{2k}(\widehat{\Upsilon}^{[2(j-k)]}+\Theta^{[j-k]})
+\widehat{\gamma}_{2k+1}(\widehat{\Upsilon}^{[2(j-k)+1]}+\Theta^{[j-k]})\right].
$$
We remark that $\widehat{\Psi}_1^{[j]}=(\widehat{\psi}_1^{[j]},\widehat{\varphi}_1^{[j]},\widehat{\chi}_1^{[j]})^T$,
and $\widehat{G}\widehat{\Psi}_1$ is the solution of System (\ref{LP}) under the plane-wave solution
(\ref{38}) and the spectral parameter (\ref{43}).

At this moment we proceed calculating
\begin{equation}
\dfrac{1}{\widehat{\lambda}_1(\epsilon)-\widehat{\lambda}_1(\epsilon)^{*}}
=\sum_{j=0}^{\infty}\sum_{m=0}^{j}\left(
\begin{array}{c}
j \\
m \\
\end{array}
\right)\dfrac{1}{2{\rm i}\varrho^{j}\sqrt{\varrho}}\epsilon^{*2(j-m)}\epsilon^{2m},
\end{equation}
and
\begin{equation}
\widehat{\Psi}_1(\epsilon)^{\dag}\widehat{\Psi}_1(\epsilon)
=\sum_{j=0}^{\infty}\sum_{m=0}^{j}\widehat{\Psi}_1^{[j-m]\dag}
\widehat{\Psi}_1^{[m]}\epsilon^{*2(j-m)}\epsilon^{2m},
\end{equation}
then it holds that
$$
\begin{array}{l}
\dfrac{\widehat{\Psi}_1(\epsilon)^{\dag}\widehat{\Psi}_1(\epsilon)}
{\widehat{\lambda}_1(\epsilon)-\widehat{\lambda}_1(\epsilon)^{*}}=
\displaystyle\sum_{j=0}^{\infty}\sum_{l=0}^{j}\left[\sum_{m=0}^{l}
\left(
\begin{array}{c}
l \\
m \\
\end{array}
\right)\dfrac{1}{2{\rm i}\varrho^{l}\sqrt{\varrho}} \epsilon^{*2(l-m)}\epsilon^{2m}\right]
\left[\sum_{n=0}^{j-l}\widehat{\Psi}_1^{[j-l-n]\dag}\widehat{\Psi}_1^{[n]}\epsilon^{*2(j-l-n)}\epsilon^{2n}\right]\\
\hspace{2.7cm}=\displaystyle\sum_{j=0}^{\infty}\sum_{s=0}^{j}\left[\sum_{l=0}^{j}
\sum_{m+n=s}^{\mbox{\tiny$\begin{array}{c}
0\leq m\leq l,\\
0\leq n\leq j-l\\
\end{array}$}}
\left(
\begin{array}{c}
l \\
m \\
\end{array}
\right)\dfrac{1}{2{\rm i}\varrho^{l}\sqrt{\varrho}}
\widehat{\Psi}_1^{[j-l-n]\dag}\widehat{\Psi}_1^{[n]}\right]\epsilon^{*2(j-s)}\epsilon^{2s}.
\end{array}$$
At present, we conclude that
\begin{equation}\label{56}
\dfrac{\widehat{\Psi}_1(\epsilon)^{\dag}\widehat{\Psi}_1(\epsilon)}
{\widehat{\lambda}_1(\epsilon)-\widehat{\lambda}_1(\epsilon)^{*}}=
\sum_{r,t=1}^{\infty}\widehat{H}_1^{[r,t]}\epsilon^{*2(r-1)}\epsilon^{2(t-1)},
\end{equation}
where
$$
\widehat{H}_1^{[r,t]}=\sum_{l=0}^{r+t-2}
\sum_{m+n=t-1}^{\mbox{\tiny$\begin{array}{c}
0\leq m\leq l,\\
0\leq n\leq r+t-l-2\\
\end{array}$}}
\left(
\begin{array}{c}
l \\
m \\
\end{array}
\right)\dfrac{1}{2{\rm i}\varrho^{l}\sqrt{\varrho}}
\widehat{\Psi}_1^{[r+t-l-n-2]\dag}\widehat{\Psi}_1^{[n]}.
$$

\subsection{The $n$th-order semirational solution}

Similarly, by Eqs. (\ref{53}) and (\ref{56}), we obtain the $n$th-order semirational solution in the form
\begin{subequations}\label{57}
\begin{align}
&\widehat{E}_{1}[n]=\left(c_1+2{\rm i}\dfrac{1}{|\widehat{H}_1|}\left|
                                \begin{array}{cc}
                                  \widehat{H}_1 & \widehat{Y}_1^{\dag} \\
                                  \widehat{X}_1 & 0 \\
                                \end{array}
                              \right|\right){\rm e}^{{\rm i}\theta_1},\\
&\widehat{E}_{2}[n]=\left(c_2+2{\rm i}\dfrac{1}{|\widehat{H}_1|}\left|
                                \begin{array}{cc}
                                  \widehat{H}_1 & \widehat{Z}_1^{\dag}\\
                                  \widehat{X}_1 & 0 \\
                                \end{array}
                              \right|\right){\rm e}^{{\rm i}\theta_1},\\
&\widehat{N}[n]=\widehat{N}_0-2{\rm i}\dfrac{\partial}{\partial t}\dfrac{1}{|\widehat{H}_1|}\left|
                                \begin{array}{cc}
                                  \widehat{H}_1 & \widehat{X}_1^{\dag} \\
                                  \widehat{X}_1 & 0 \\
                                \end{array}
                              \right|,\\
&\widehat{p}_{1}[n]={\rm i}c_1b_1{\rm e}^{{\rm i}\theta_1}+2{\rm i}\dfrac{\partial}{\partial t}\left(\dfrac{1}{|\widehat{H}_1|}\left|
                                \begin{array}{cc}
                                  \widehat{H}_1 & \widehat{Y}_1^{\dag} \\
                                  \widehat{X}_1 & 0 \\
                                \end{array}
                              \right|{\rm e}^{{\rm i}\theta_1}\right),\\
&\widehat{p}_{2}[n]={\rm i}c_2b_1{\rm e}^{{\rm i}\theta_1}+2{\rm i}\dfrac{\partial}{\partial t}\left(\dfrac{1}{|\widehat{H}_1|}\left|
                                \begin{array}{cc}
                                  \widehat{H}_1 & \widehat{Z}_1^{\dag} \\
                                  \widehat{X}_1 & 0 \\
                                \end{array}
                              \right|{\rm e}^{{\rm i}\theta_1}\right),\\
&\widehat{M}_{11}[n]=-2a_1b_1-4b_1\omega-N_0+2{\rm i}\dfrac{\partial}{\partial t}\left(\dfrac{1}{|\widehat{H}_1|}\left|
                                \begin{array}{cc}
                                  \widehat{H}_1 & \widehat{Y}_1^{\dag} \\
                                  \widehat{Y}_1 & 0 \\
                                \end{array}
                               \right|\right),\\
&\widehat{M}_{12}[n]=2{\rm i}\dfrac{\partial}{\partial t}\left(\dfrac{1}{|\widehat{H}_1|}\left|
                                \begin{array}{cc}
                                  \widehat{H}_1 & \widehat{Z}_1^{\dag} \\
                                  \widehat{Y}_1 & 0 \\
                                \end{array}
                              \right|\right),\\
&\widehat{M}_{21}[n]=2{\rm i}\dfrac{\partial}{\partial t}\left(\dfrac{1}{|\widehat{H}_1|}\left|
                                \begin{array}{cc}
                                  \widehat{H}_1 & \widehat{Y}_1^{\dag} \\
                                  \widehat{Z}_1 & 0 \\
                                \end{array}
                              \right|\right),\\
&\widehat{M}_{22}[n]=-2a_1b_1-4b_1\omega-N_0+2{\rm i}\dfrac{\partial}{\partial t}\left(\dfrac{1}{|\widehat{H}_1|}\left|
                                \begin{array}{cc}
                                  \widehat{H}_1 & \widehat{Z}_1^{\dag} \\
                                  \widehat{Z}_1 & 0 \\
                                \end{array}
                              \right|\right),
\end{align}
\end{subequations}
where
$$\begin{array}{l}
\widehat{H}_{1}=\left(\widehat{H}_1^{[r,t]}\right)_{1\leq r,t\leq n},\\
\widehat{X}_1=(\widehat{\psi}_1^{[0]},\widehat{\psi}_1^{[1]},\cdots,\widehat{\psi}_1^{[n-1]}),\\
\widehat{Y}_1=(\widehat{\varphi}_1^{[0]},\widehat{\varphi}_1^{[1]},\cdots,\widehat{\varphi}_1^{[n-1]}),\\
\widehat{Z}_1=(\widehat{\chi}_1^{[0]},\widehat{\chi}_1^{[1]},\cdots,\widehat{\chi}_1^{[n-1]}).
\end{array}$$
Specifically, the first-order semirational rogue wave solution holds
\begin{subequations}\label{}
\begin{align}
&\widehat{E}_{1}[1]=\left(c_1-2{\rm i}\dfrac{\widehat{\psi}_1^{[0]}\widehat{\varphi}_1^{[0]*}}{\widehat{H}_1^{[1,1]}}\right){\rm e}^{{\rm i}\theta_1},\\
&\widehat{E}_{2}[1]=\left(c_2-2{\rm i}\dfrac{\widehat{\psi}_1^{[0]}\widehat{\chi}_1^{[0]*}}{\widehat{H}_1^{[1,1]}}\right){\rm e}^{{\rm i}\theta_1},\\
&\widehat{N}[1]=N_0+2{\rm i}\dfrac{\widehat{\psi}_1^{[0]}\widehat{\psi}_1^{[0]*}}{\widehat{H}_1^{[1,1]}},\\
&\widehat{p}_{1}[1]={\rm i}c_1b_1{\rm e}^{{\rm i}\theta_1}-2{\rm i}\dfrac{\partial}{\partial t}\left(\dfrac{\widehat{\psi}_1^{[0]}\widehat{\varphi}_1^{[0]*}}{\widehat{H}_1^{[1,1]}}{\rm e}^{{\rm i}\theta_1}\right),\\
&\widehat{p}_{2}[1]={\rm i}c_2b_1{\rm e}^{{\rm i}\theta_1}-2{\rm i}\dfrac{\partial}{\partial t}\left(\dfrac{\widehat{\psi}_1^{[0]}\widehat{\chi}_1^{[0]*}}{\widehat{H}_1^{[1,1]}}{\rm e}^{{\rm i}\theta_1}\right),\\
&\widehat{M}_{11}[1]=-2a_1b_1-4b_1\omega-N_0-2{\rm i}\dfrac{\partial}{\partial t}\left(\dfrac{\widehat{\varphi}_1^{[0]}
\widehat{\varphi}_1^{[0]*}}{\widehat{H}_1^{[1,1]}}\right),\\
&\widehat{M}_{12}[1]=-2{\rm i}\dfrac{\partial}{\partial t}\left(\dfrac{\widehat{\varphi}_1^{[0]}\widehat{\chi}_1^{[0]*}}{\widehat{H}_1^{[1,1]}}\right),\\
&\widehat{M}_{21}[1]=-2{\rm i}\dfrac{\partial}{\partial t}\left(\dfrac{\widehat{\chi}_1^{[0]}\widehat{\varphi}_1^{[0]*}}{\widehat{H}_1^{[1,1]}}\right),\\
&\widehat{M}_{22}[1]=-2a_1b_1-4b_1\omega-N_0-2{\rm i}\dfrac{\partial}{\partial t}\left(\dfrac{\widehat{\chi}_1^{[0]}\widehat{\chi}_1^{[0]*}}{\widehat{H}_1^{[1,1]}}\right).
\end{align}
\end{subequations}
and the second-order semirational rogue wave solution is given by
\begin{subequations}\label{}
\begin{align}
&\widehat{E}_{1}[2]=\left[c_1+2{\rm i}\dfrac{\widehat{\varphi}_1^{[0]*}(\widehat{\psi}_1^{[1]}\widehat{H}_1^{[2,1]}-\widehat{\psi}_1^{[0]}\widehat{H}_1^{[2,2]})
+\widehat{\varphi}_1^{[1]*}(\widehat{\psi}_1^{[0]}\widehat{H}_1^{[1,2]}-\widehat{\psi}_1^{[1]}\widehat{H}_1^{[1,1]})}
{\widehat{H}_1^{[1,1]}\widehat{H}_1^{[2,2]}-\widehat{H}_1^{[1,2]}\widehat{H}_1^{[2,1]}}\right]{\rm e}^{{\rm i}\theta_1},\\
&\widehat{E}_{2}[2]=\left[c_2+2{\rm i}\dfrac{\widehat{\chi}_1^{[0]*}(\widehat{\psi}_1^{[1]}\widehat{H}_1^{[2,1]}-\widehat{\psi}_1^{[0]}\widehat{H}_1^{[2,2]})+
\widehat{\chi}_1^{[1]*}(\widehat{\psi}_1^{[0]}\widehat{H}_1^{[1,2]}-\widehat{\psi}_1^{[1]}\widehat{H}_1^{[1,1]})}
{\widehat{H}_1^{[1,1]}\widehat{H}_1^{[2,2]}-\widehat{H}_1^{[1,2]}\widehat{H}_1^{[2,1]}}\right]{\rm e}^{{\rm i}\theta_1},\\
&\widehat{N}[2]=N_0-2{\rm i}\dfrac{
\widehat{\psi}_1^{[0]*}(\widehat{\psi}_1^{[1]}\widehat{H}_1^{[2,1]}-\widehat{\psi}_1^{[0]}\widehat{H}_1^{[2,2]})
+\widehat{\psi}_1^{[1]*}(\widehat{\psi}_1^{[0]}\widehat{H}_1^{[1,2]}-\widehat{\psi}_1^{[1]}\widehat{H}_1^{[1,1]})}{\widehat{H}_1^{[1,1]}\widehat{H}_1^{[2,2]}-
\widehat{H}_1^{[1,2]}\widehat{H}_1^{[2,1]}},\\
&\widehat{p}_{1}[2]={\rm i}c_1b_1{\rm e}^{{\rm i}\theta_1}+2{\rm i}\dfrac{\partial}{\partial t}\left[\dfrac
{\widehat{\varphi}_1^{[0]*}(\widehat{\psi}_1^{[1]}\widehat{H}_1^{[2,1]}-\widehat{\psi}_1^{[0]}\widehat{H}_1^{[2,2]})
+\widehat{\varphi}_1^{[1]*}(\widehat{\psi}_1^{[0]}\widehat{H}_1^{[1,2]}-\widehat{\psi}_1^{[1]}\widehat{H}_1^{[1,1]})}
{\widehat{H}_1^{[1,1]}\widehat{H}_1^{[2,2]}-\widehat{H}_1^{[1,2]}\widehat{H}_1^{[2,1]}}{\rm e}^{{\rm i}\theta_1}\right],\\
&\widehat{p}_{2}[2]={\rm i}c_2b_1{\rm e}^{{\rm i}\theta_1}+2{\rm i}\dfrac{\partial}{\partial t}\left[\dfrac{
\widehat{\chi}_1^{[0]*}(\widehat{\psi}_1^{[1]}\widehat{H}_1^{[2,1]}-\widehat{\psi}_1^{[0]}\widehat{H}_1^{[2,2]})
+\widehat{\chi}_1^{[1]*}(\widehat{\psi}_1^{[0]}\widehat{H}_1^{[1,2]}-\widehat{\psi}_1^{[1]}\widehat{H}_1^{[1,1]})}
{\widehat{H}_1^{[1,1]}\widehat{H}_1^{[2,2]}-\widehat{H}_1^{[1,2]}\widehat{H}_1^{[2,1]}}{\rm e}^{{\rm i}\theta_1}\right],\\
&\widehat{M}_{11}[2]=-2a_1b_1-4b_1\omega-N_0+2{\rm i}\dfrac{\partial}{\partial t}\left[\dfrac
{\widehat{\varphi}_1^{[0]*}(\widehat{\varphi}_1^{[1]}\widehat{H}_1^{[2,1]}-\widehat{\varphi}_1^{[0]}\widehat{H}_1^{[2,2]})
+\widehat{\varphi}_1^{[1]*}(\widehat{\varphi}_1^{[0]}\widehat{H}_1^{[1,2]}-\widehat{\varphi}_1^{[1]}\widehat{H}_1^{[1,1]})}
{\widehat{H}_1^{[1,1]}\widehat{H}_1^{[2,2]}-
\widehat{H}_1^{[1,2]}\widehat{H}_1^{[2,1]}}\right],\\
&\widehat{M}_{12}[2]=2{\rm i}\dfrac{\partial}{\partial t}\left[\dfrac{\widehat{\chi}_1^{[0]*}(\widehat{\varphi}_1^{[1]}\widehat{H}_1^{[2,1]}
-\widehat{\varphi}_1^{[0]}\widehat{H}_1^{[2,2]})
+\widehat{\chi}_1^{[1]*}(\widehat{\varphi}_1^{[0]}H_1^{[1,2]}
-\widehat{\varphi}_1^{[1]}\widehat{H}_1^{[1,1]})}{\widehat{H}_1^{[1,1]}\widehat{H}_1^{[2,2]}-
\widehat{H}_1^{[1,2]}\widehat{H}_1^{[2,1]}}\right],\\
&\widehat{M}_{21}[2]=2{\rm i}\dfrac{\partial}{\partial t}\left[\dfrac{\widehat{\varphi}_1^{[0]*}(\widehat{\chi}_1^{[1]}\widehat{H}_1^{[2,1]}
-\widehat{\chi}_1^{[0]}\widehat{H}_1^{[2,2]})
+\widehat{\varphi}_1^{[1]*}(\widehat{\chi}_1^{[0]}\widehat{H}_1^{[1,2]}
-\widehat{\chi}_1^{[1]}\widehat{H}_1^{[1,1]})}{\widehat{H}_1^{[1,1]}\widehat{H}_1^{[2,2]}-
\widehat{H}_1^{[1,2]}\widehat{H}_1^{[2,1]}}\right],\\
&\widehat{M}_{22}[2]=-2a_1b_1-4b_1\omega-N_0+2{\rm i}\dfrac{\partial}{\partial t}\left[\dfrac{
\widehat{\chi}_1^{[0]*}(\widehat{\chi}_1^{[1]}\widehat{H}_1^{[2,1]}-\widehat{\chi}_1^{[0]}\widehat{H}_1^{[2,2]})
+\widehat{\chi}_1^{[1]*}(\widehat{\chi}_1^{[0]}\widehat{H}_1^{[1,2]}-\widehat{\chi}_1^{[1]}\widehat{H}_1^{[1,1]})}{\widehat{H}_1^{[1,1]}\widehat{H}_1^{[2,2]}-
\widehat{H}_1^{[1,2]}\widehat{H}_1^{[2,1]}}\right],
\end{align}
\end{subequations}
where
$$
\begin{array}{l}
\widehat{H}_1^{[1,1]}=\dfrac{\widehat{\Psi}_1^{[0]\dag}\widehat{\Psi}_1^{[0]}}{2{\rm i}\sqrt{\varrho}},
\ \widehat{H}_1^{[1,2]}=\dfrac{1}{2{\rm i}\sqrt{\varrho}}\left[\widehat{\Psi}_1^{[0]\dag}\widehat{\Psi}_1^{[1]}
+\dfrac{\widehat{\Psi}_1^{[0]\dag}\widehat{\Psi}_1^{[0]}}{\varrho}\right],\\
\widehat{H}_1^{[2,1]}=\dfrac{1}{2{\rm i}\sqrt{\varrho}}\left[\widehat{\Psi}_1^{[1]\dag}\widehat{\Psi}_1^{[0]}
+\dfrac{\widehat{\Psi}_1^{[0]\dag}\widehat{\Psi}_1^{[0]}}{\varrho}\right],\\
\widehat{H}_1^{[2,2]}=\dfrac{1}{2{\rm i}\sqrt{\varrho}}
\left[\widehat{\Psi}_1^{[1]\dag}\widehat{\Psi}_1^{[1]}
+\dfrac{\widehat{\Psi}_1^{[1]\dag}\widehat{\Psi}_1^{[0]}\varrho
+\widehat{\Psi}_1^{[0]\dag}\widehat{\Psi}_1^{[0]}}{\varrho^2}+
\dfrac{\widehat{\Psi}_1^{[0]\dag}\widehat{\Psi}_1^{[1]}\varrho
+\widehat{\Psi}_1^{[0]\dag}\widehat{\Psi}_1^{[0]}}{\varrho^2}
\right].
\end{array}$$

As an example, we choose the specific parameters
$$
c_1=1,c_2=0,a_1=0,m_1=m_2=1,N_0=1,\widehat{\gamma}_0=1,\widehat{\gamma}_1=10,b_1=-\dfrac{1}{2},
\omega=1,\alpha=\dfrac{1}{5000},
$$
then we can present the interactions between solitons and rogue waves with different types, which can be
summarize as
\begin{center}
Table 3: types of interactions between solitons and rogue waves in each component.
\begin{tabular}{lcccccccc}
\hline
component &$E_1$ &$E_2$ &$N$ &$p_1$ &$p_2$ &$M_{11}$ &$M_{12}$/$M_{21}$ &$M_{22}$ \\
\hline
$\begin{array}{l}{\rm soliton+}\\{\rm rogue~wave} \end{array}$
& $\begin{array}{l}{\rm dark+}\\{\rm bright} \end{array}$
& $\begin{array}{l}{\rm bright+}\\{\rm bright} \end{array}$
& $\begin{array}{l}{\rm dark+}\\{\rm dark} \end{array}$
& $\begin{array}{l}{\rm dark+}\\{\rm bright} \end{array}$
& $\begin{array}{l}{\rm bright+}\\{\rm bright} \end{array}$
& $\begin{array}{l}{\rm bright+}\\{\rm dark} \end{array}$
&$\begin{array}{l}{\rm bright+}\\{\rm bright} \end{array}$
&$\begin{array}{l}{\rm dark+}\\{\rm bright} \end{array}$ \\
\hline
\end{tabular}
\end{center}

\begin{figure}[!h]
\centering
\renewcommand{\figurename}{{\bf Fig.}}
{\includegraphics[height=4cm,width=6cm]{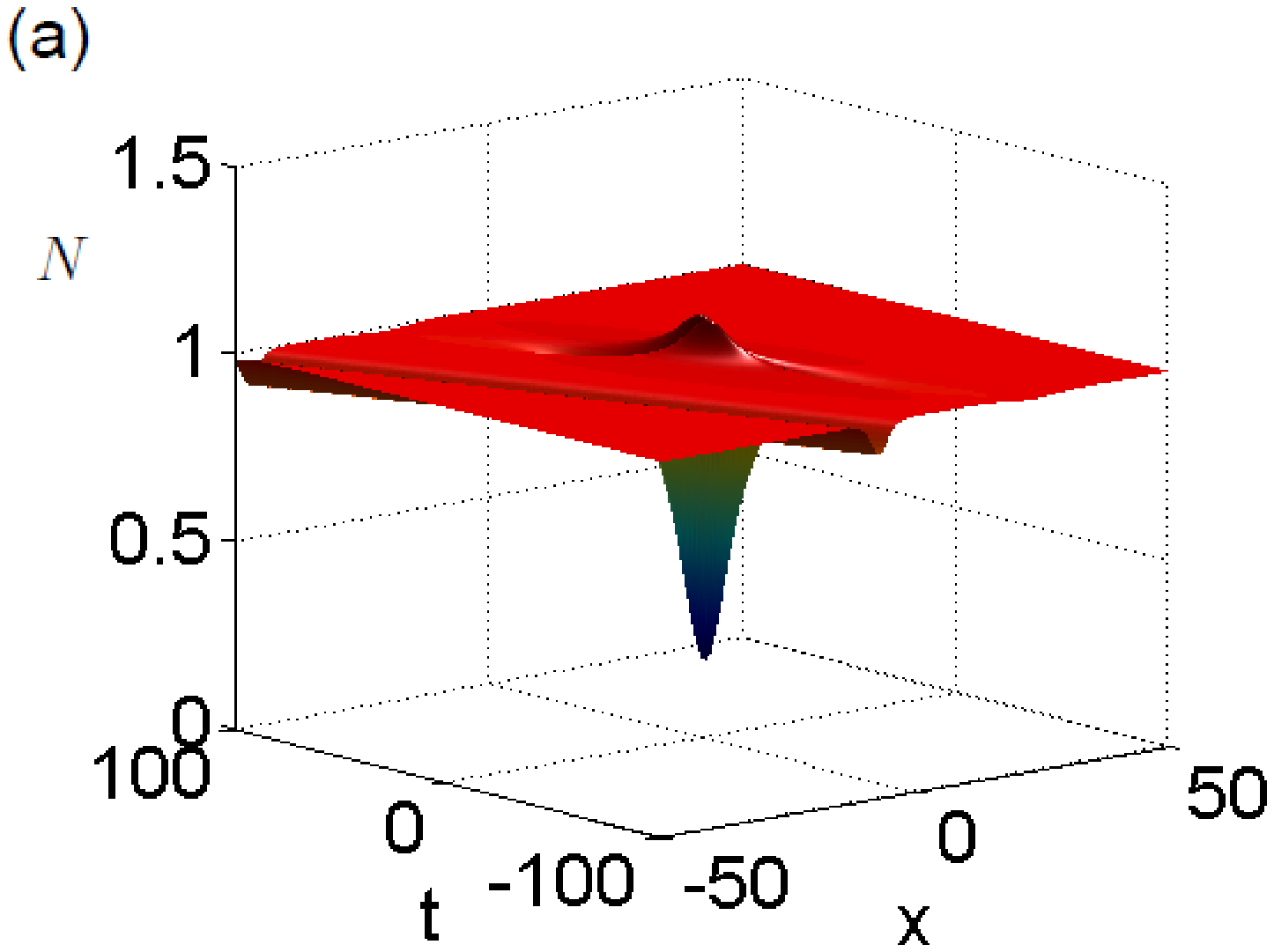}}
{\includegraphics[height=4cm,width=6cm]{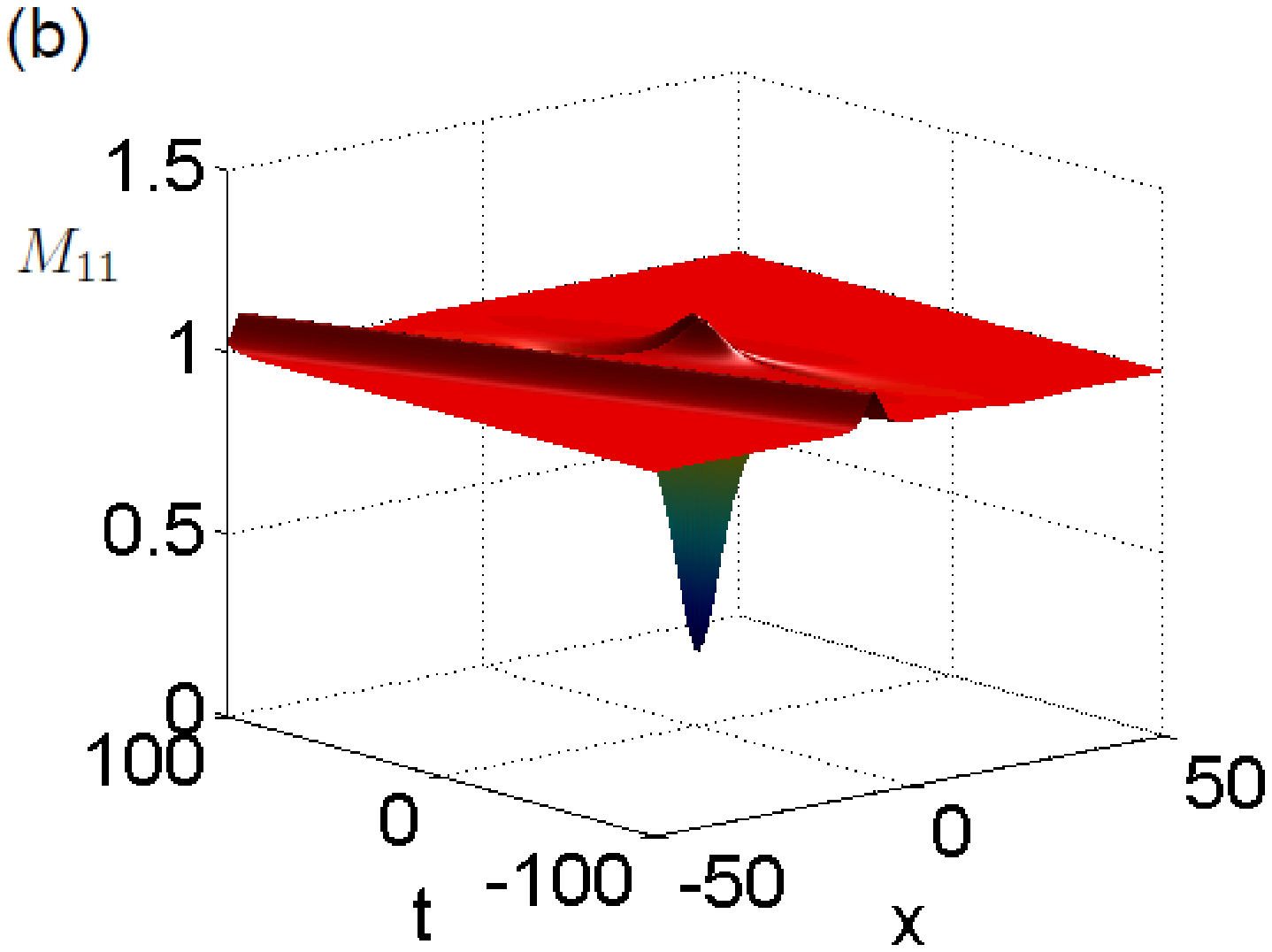}}
\caption{(a), (b) Evolution plots of the interaction between a dark soliton and a dark rogue wave
in the $N$ component and the interaction between a bright soliton and a dark rogue wave in the $M_{11}$ component, respectively. }
\label{fig:5}
\end{figure}

Unlike the known dark-bright and bright-bright interactions between solitons and rogue waves
in the coupled NLS equations \cite{37} and the coupled Hirota equations \cite{38},
here we show some novel interactional structures in the CMB equations.
For example, Fig. \ref{fig:5}(a) displays the interaction between a dark soliton and a dark rogue wave in the
$N$ component, and Fig. \ref{fig:5}(b) exhibits the interaction between a bright soliton
and a dark rogue wave in the $M_{11}$ component.

\begin{figure}[!h]
\centering
\renewcommand{\figurename}{{\bf Fig.}}
{\includegraphics[height=4cm,width=6cm]{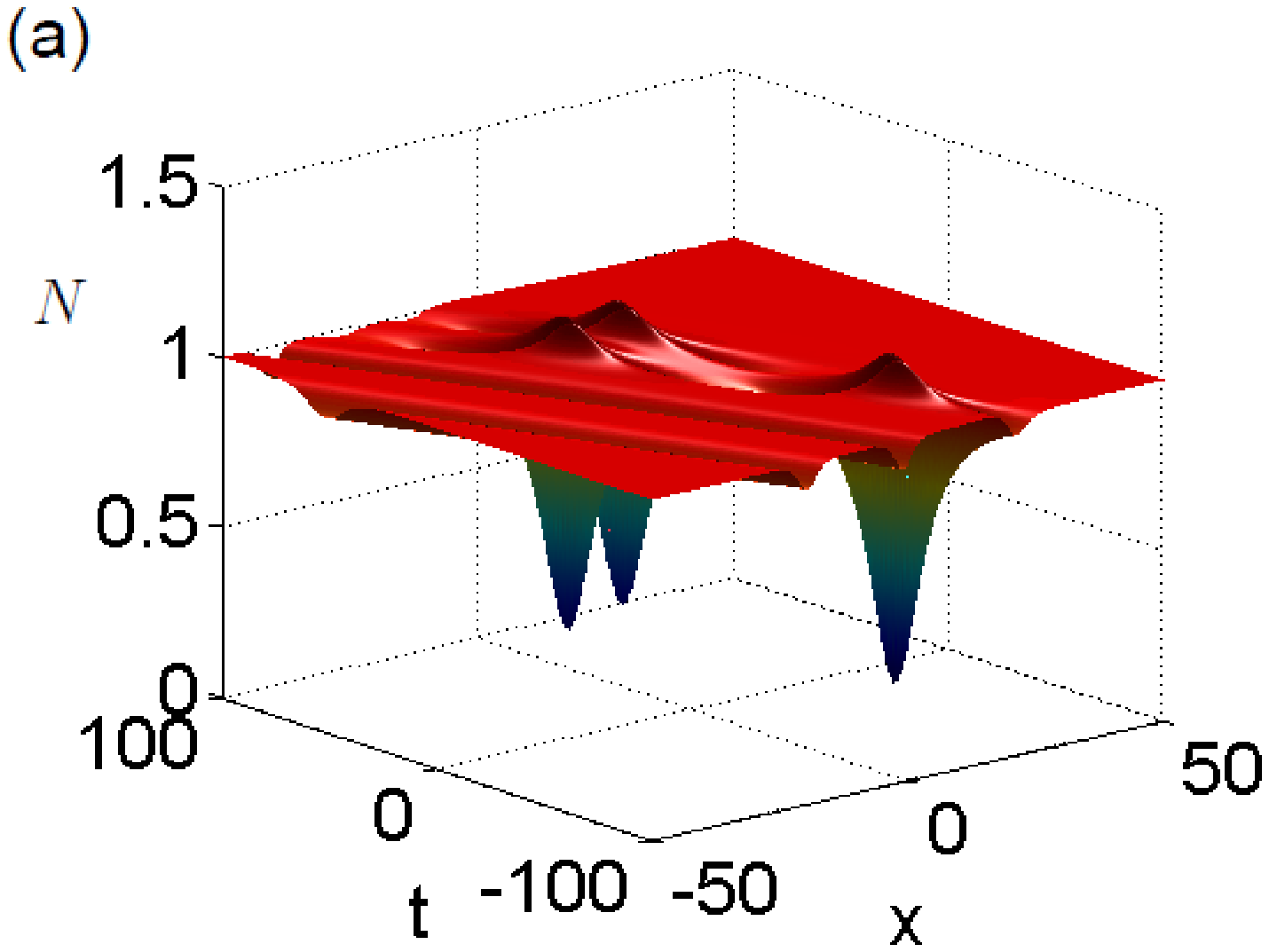}}
{\includegraphics[height=4cm,width=6cm]{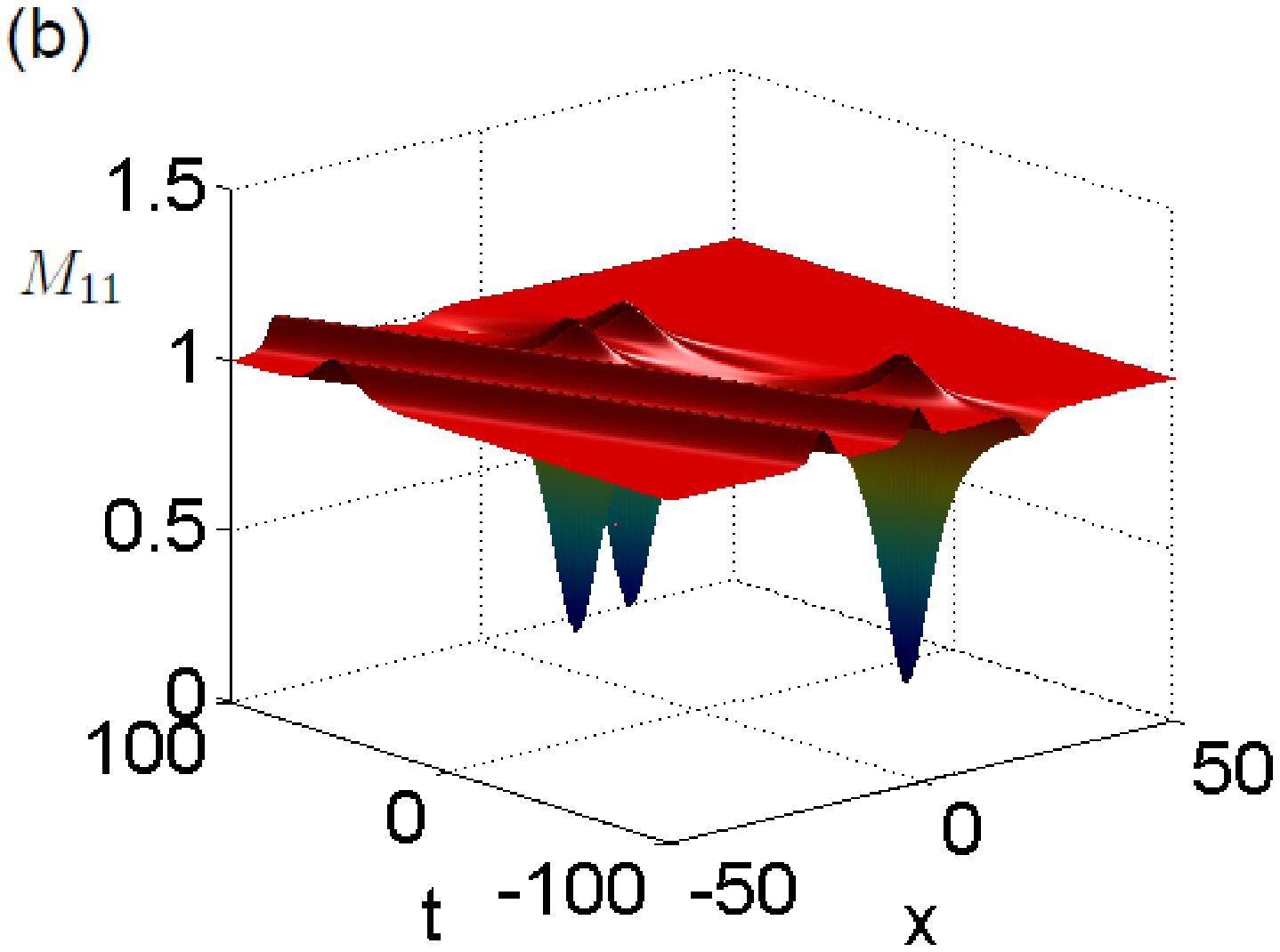}}
\caption{(a), (b) Evolution plots of the interaction between two dark solitons and triplet dark rogue waves
in the $N$ component and the interaction between two bright solitons and triplet dark rogue waves in the $M_{11}$ component, respectively.}
\label{fig:6}
\end{figure}

\begin{figure}[!h]
\centering
\renewcommand{\figurename}{{\bf Fig.}}
{\includegraphics[height=4cm,width=6cm]{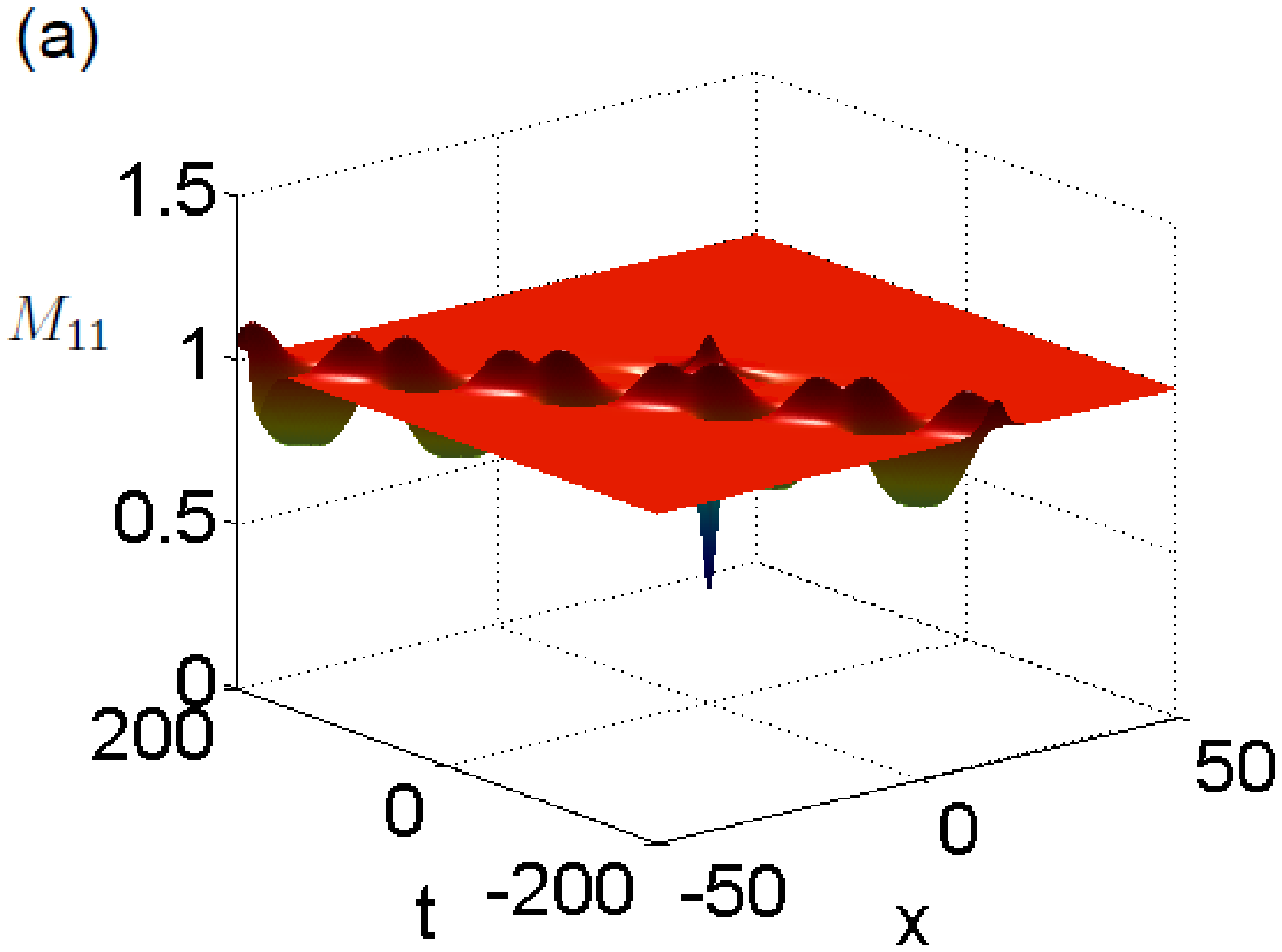}}
{\includegraphics[height=4cm,width=6cm]{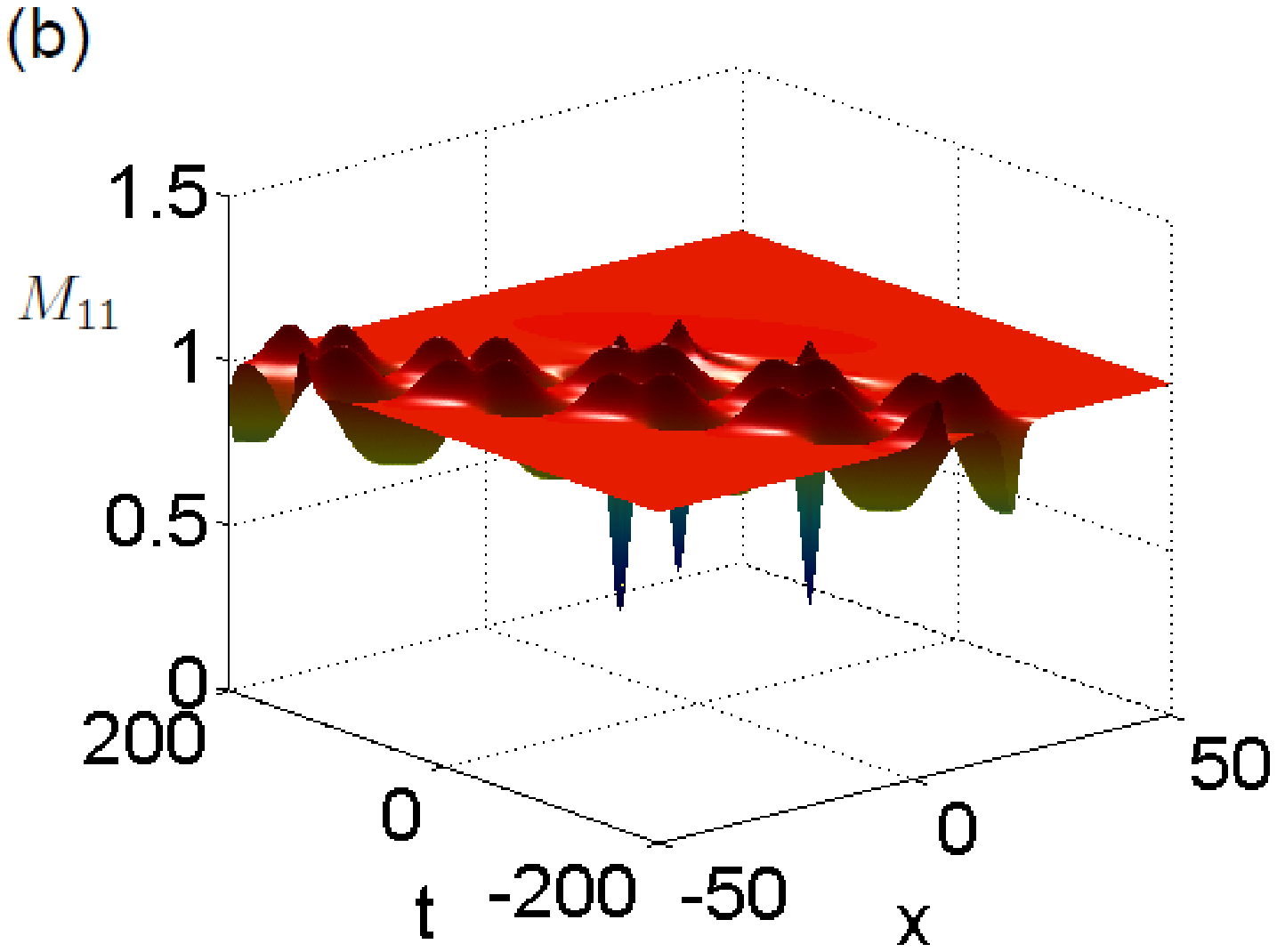}}
\caption{Evolution plots of the interactions between breathers and dark rogue waves
in the $M_{11}$ component. The parameters are same as Figs. \ref{fig:5} and \ref{fig:6}
except $c_1=c_2=1$.}
\label{fig:7}
\end{figure}

Noteworthily, we should say that the free parameter $\alpha$ controls the separation and
mergence between solitons and rogue waves \cite{38}. When increasing the value of $\alpha$, the soliton and the
rogue wave can merge with each other.  Next by choosing
$\widehat{\gamma}_0=0,\widehat{\gamma}_1=1,\widehat{\gamma}_2=100,\widehat{\gamma}_3=0,$
we can demonstrate the interactions between
two dark/bright solitons and the second-order dark rogue wave with triplet pattern,
see Fig. \ref{fig:6}. In adition, the interactions between breathers and dark rogue waves
can also be obtained by choosing the above specific parameters except
$c_1=c_2=1$, see Fig. \ref{fig:7}.

\section{Modulation instability and W-shaped solitons}

In this section, we perform the standard linearized stability analysis on the plane-wave solution (\ref{08}).
A perturbed plane-wave solution can be given by
\begin{align}
&E_1=E_1[0][1+r_1{\rm e}^{{\rm i}\kappa(x-\Omega t)}+r_2^{*}{\rm e}^{-{\rm i}\kappa(x-\Omega^{*}t)}],\nonumber\\
&E_2=E_2[0][1+r_3{\rm e}^{{\rm i}\kappa(x-\Omega t)}+r_4^{*}{\rm e}^{-{\rm i}\kappa(x-\Omega^{*}t)}],\nonumber\\
&N=N[0]+r_5{\rm e}^{{\rm i}\kappa(x-\Omega t)}+r_5^{*}{\rm e}^{-{\rm i}\kappa(x-\Omega^{*}t)},\nonumber\\
&p_1=E_{1t},\ p_2=E_{2t},\label{60}\\
&M_{11}=M_{11}[0]+r_6{\rm e}^{{\rm i}\kappa(x-\Omega t)}+r_6^{*}{\rm e}^{-{\rm i}\kappa(x-\Omega^{*}t)},\nonumber\\
&M_{12}=M_{21}^{*}=M_{12}[0][r_7{\rm e}^{{\rm i}\kappa(x-\Omega t)}+r_8^{*}{\rm e}^{-{\rm i}\kappa(x-\Omega^{*}t)}],\nonumber\\
&M_{22}=M_{22}[0]+r_9{\rm e}^{{\rm i}\kappa(x-\Omega t)}+r_9^{*}{\rm e}^{-{\rm i}\kappa(x-\Omega^{*}t)},\nonumber
\end{align}
where $\kappa$ and $\Omega$ are assumed to be real and complex, respectively.
Substituting Eq. (\ref{60}) into the CMB system (\ref{02}) and taking the linear part for
$R=(r_1,r_2,\cdots,r_9)^{T}$, one can obtain a algebraic equation
$\mathcal{M}R=0$. Here $\mathcal{M}=(\mathcal{M}_{ij})_{1\leq i,j\leq 9}$ is a $9\times9$ matrix:
$$
\mathcal{M}=\left(
  \begin{array}{ccccccccc}
    \Omega c_1^2 &\Omega c_1^2 &\Omega c_2^2 &\Omega c_2^2& -2& 0&0&0 &0\\
    \mathcal{M}_{21} &0 & c_2^2\sigma&0 &-2\delta&-2\delta &0 &c_2^2\sigma &0\\
    0& \mathcal{M}_{32}& 0& c_2^2\sigma& -2\delta&-2\delta& c_2^2\sigma& 0& 0\\
   c_1^2\sigma& 0& \mathcal{M}_{43}&0& -2\delta& 0& c_1^2\sigma& 0& -2\delta\\
    0& c_1^2\sigma& 0& \mathcal{M}_{54}& -2\delta& 0& 0& c_1^2\sigma&-2\delta\\
    -\Omega c_1^2\kappa & -\Omega c_1^2 \kappa & 0 & 0 & 0 & 2\kappa & 0&0&0 \\
    0& -\delta(\Omega\kappa+\sigma)& -\delta(\Omega\kappa+\sigma)&0& 0& 0& \sigma(\delta-\kappa)&0&0\\
 \delta(\Omega\kappa-\sigma)& 0& 0& \delta(\Omega\kappa-\sigma)& 0& 0& 0& \sigma(\delta+\kappa)& 0\\
    0& 0& -\Omega c_2^2\kappa & -\Omega c_2^2\kappa& 0&0&0&0&2\kappa
  \end{array}
\right),
$$
with
$$\begin{array}{l}
\mathcal{M}_{21}=4\Omega\delta\kappa^2+4\delta(\Omega a_1+2\Omega\omega-b_1)\kappa-c_2^2\sigma,\\
\mathcal{M}_{32}=4\Omega\delta\kappa^2-4\delta(\Omega a_1+2\Omega\omega-b_1)\kappa-c_2^2\sigma,\\
\mathcal{M}_{43}=4\Omega\delta\kappa^2+4\delta(\Omega a_2+2\Omega\omega-b_2)\kappa-c_1^2\sigma,\\
\mathcal{M}_{54}=4\Omega\delta\kappa^2-4\delta(\Omega a_2+2\Omega\omega-b_2)\kappa-c_1^2\sigma.
\end{array}$$
As suggested in Refs. \cite{39,40,41}, the state transition between rogue wave and  W-shaped soliton
arise from the attenuation of the MI growth rate in the zero frequency region. For this study,
we solve $\det(\mathcal{M})=0$ and can get a quartic equation with respect to $\Omega$.
Since the quartic equation is very complicated, here, for simplicity,
we take $c_1=c_2=1,a_1=-a_2=-1,\omega=1$, then it follows that
\begin{equation}\label{61}
F_0\Omega^4+F_1\Omega^3+F_2\Omega^3+F_3\Omega^4+F_4=0,
\end{equation}
where
$$\begin{array}{l}
F_0=4(16\kappa^6-272\kappa^4+1156\kappa^2-1384),\\
F_1=4(88b_1\kappa^4+40 b_2\kappa^4-564 b_1\kappa^2-524 b_2\kappa^2+784 b_1+1408 b_2),\\
F_2=-64 b_1^2 \kappa^4-64 b_2^2 \kappa^4+356 b_1^2 \kappa^2+760 b_1 b_2 \kappa^2
+996 b_2^2 \kappa^2-396 b_1^2-3384 b_1 b_2-2508 b_2^2,\\
F_3=16 b_1^3 \kappa^2-144 b_1^2 b_2 \kappa^2-368 b_1 b_2^2 \kappa^2-16 b_2^3 \kappa^2
-44 b_1^3+516 b_1^2 b_2+1644 b_1 b_2^2-68 b_2^3,\\
F_4=-b_2^4-318 b_1^2 b_2^2+64 b_1^2 b_2^2 \kappa^2+32 b_1^3 b_2+32 b_1 b_2^3-b_1^4.
\end{array}$$

\begin{figure}[!h]
\centering
\renewcommand{\figurename}{{\bf Fig.}}
{\includegraphics[height=4cm,width=6cm]{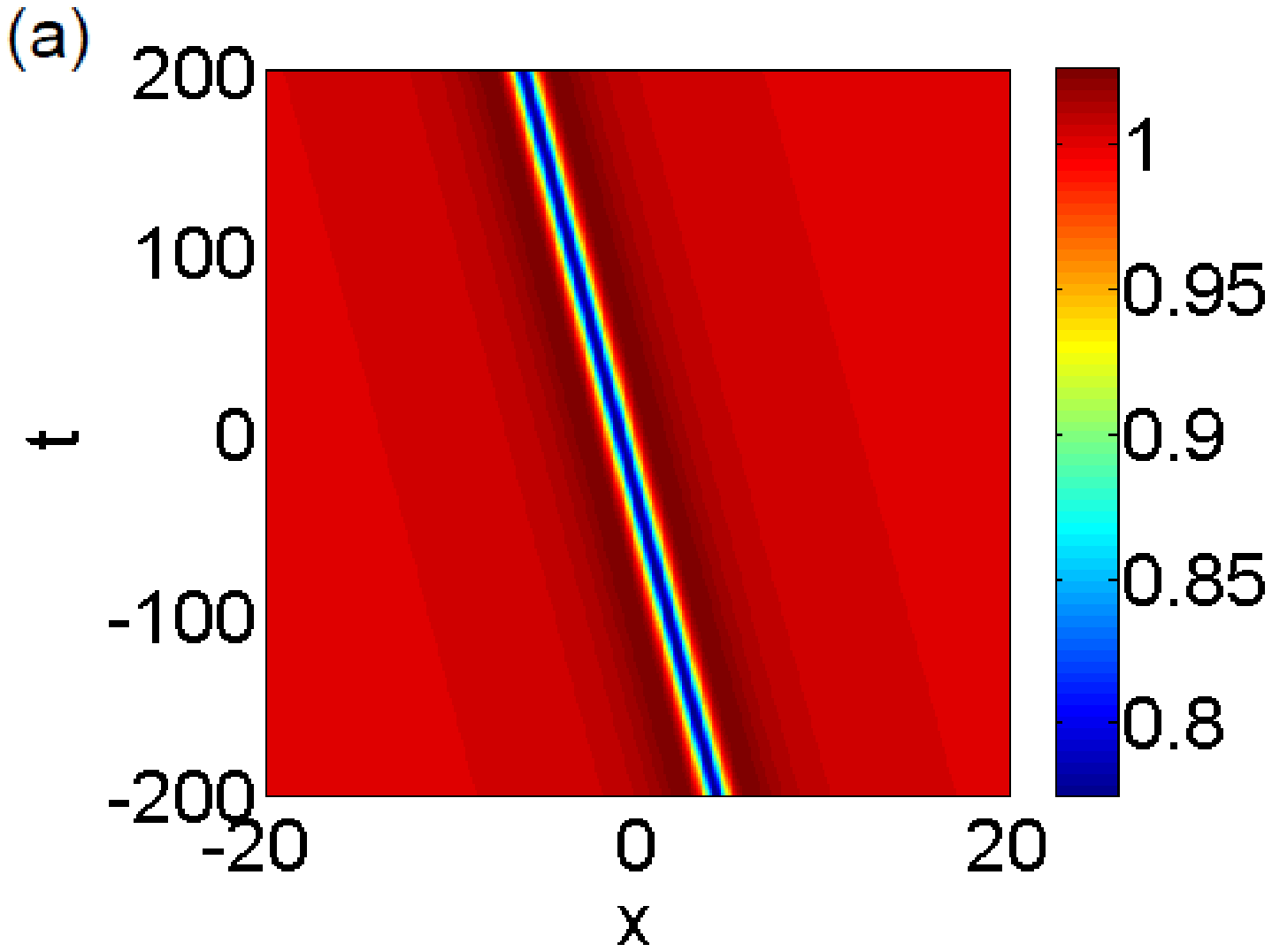}}
{\includegraphics[height=4cm,width=6cm]{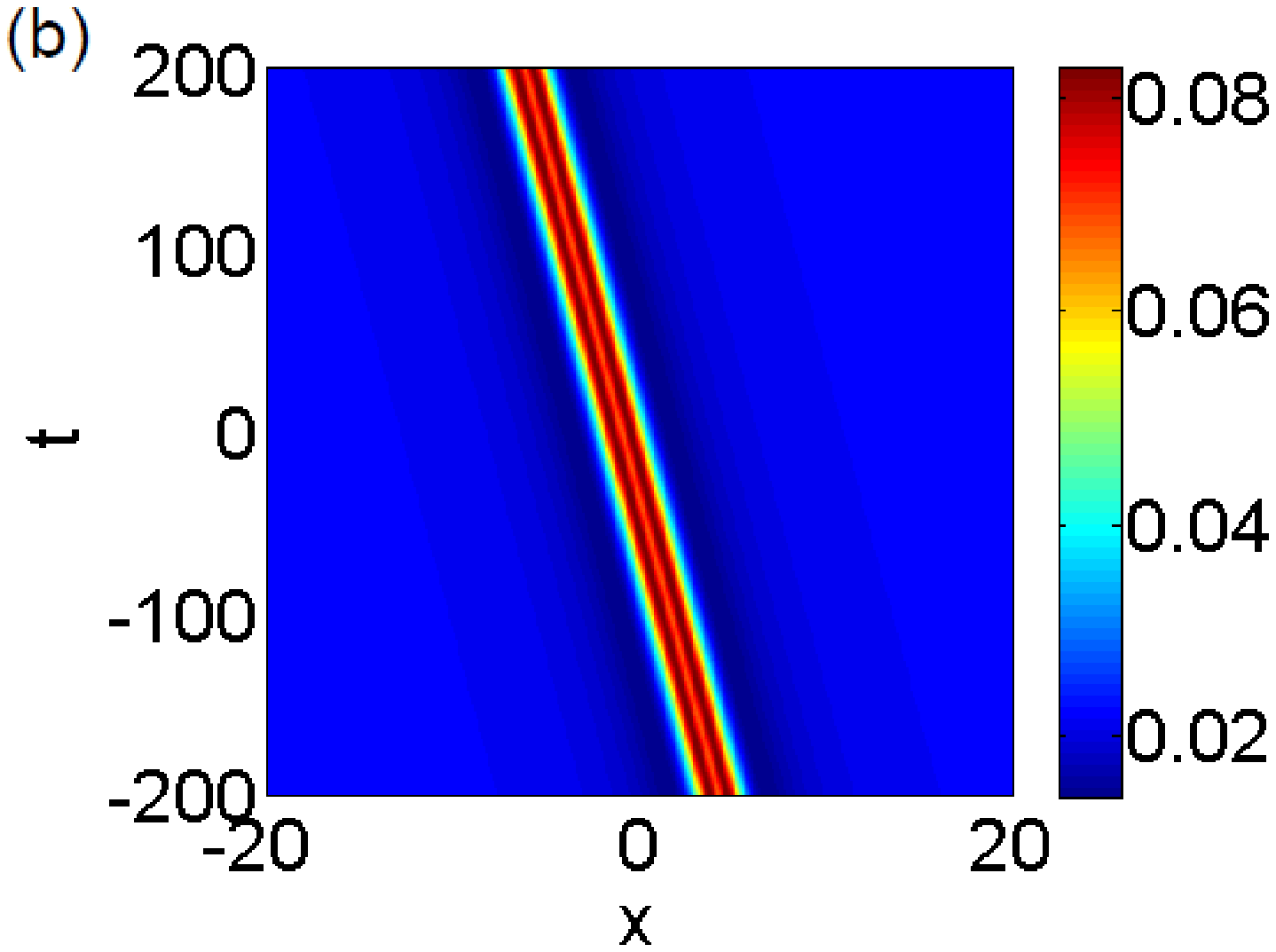}}
\caption{(a), (b) Density plots of the dark W-shaped soliton in the $E_1$ component and
the double-peak W-shaped soliton in the $p_2$ component, respectively.}
\label{fig:8}
\end{figure}

\begin{figure}[!h]
\centering
\renewcommand{\figurename}{{\bf Fig.}}
{\includegraphics[height=4cm,width=6cm]{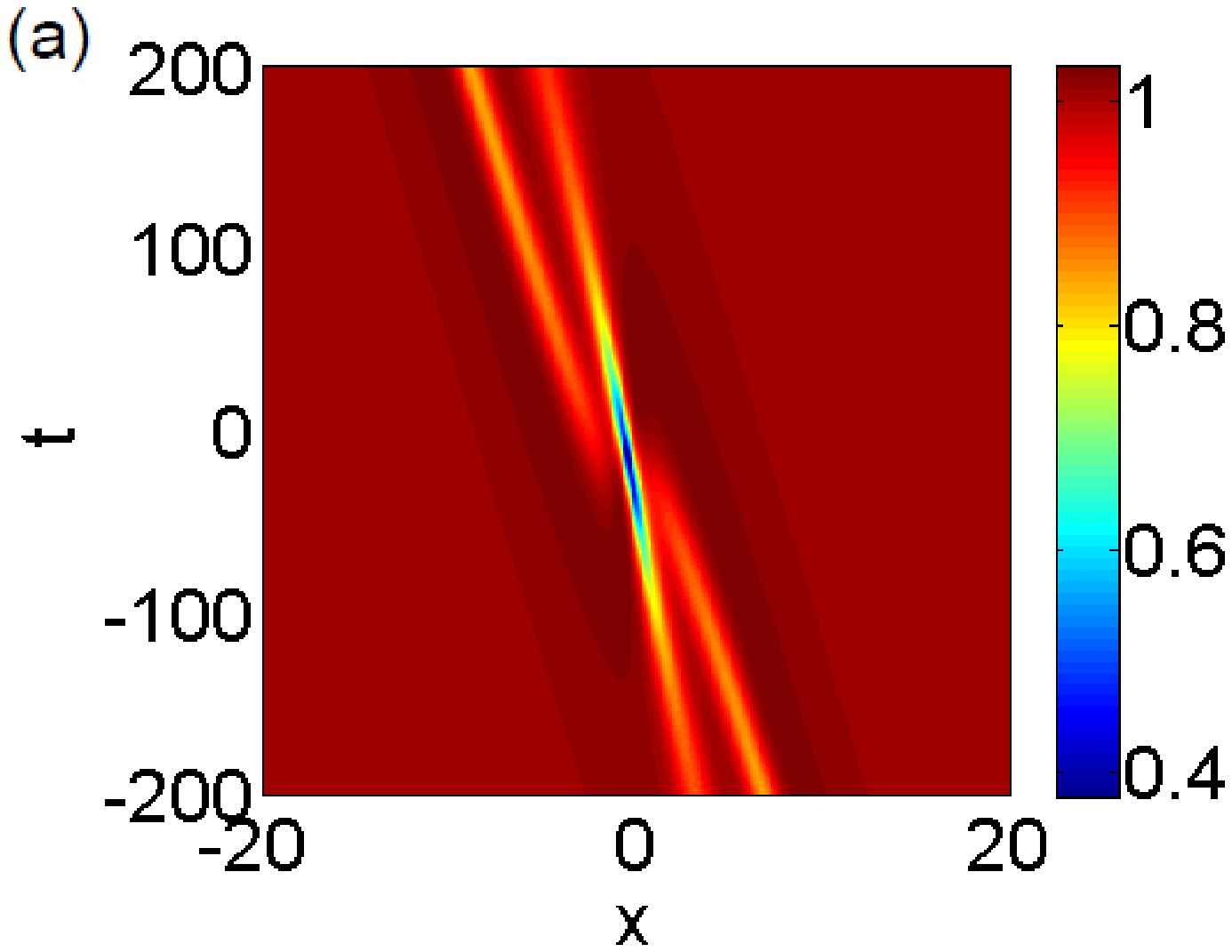}}
{\includegraphics[height=4cm,width=6cm]{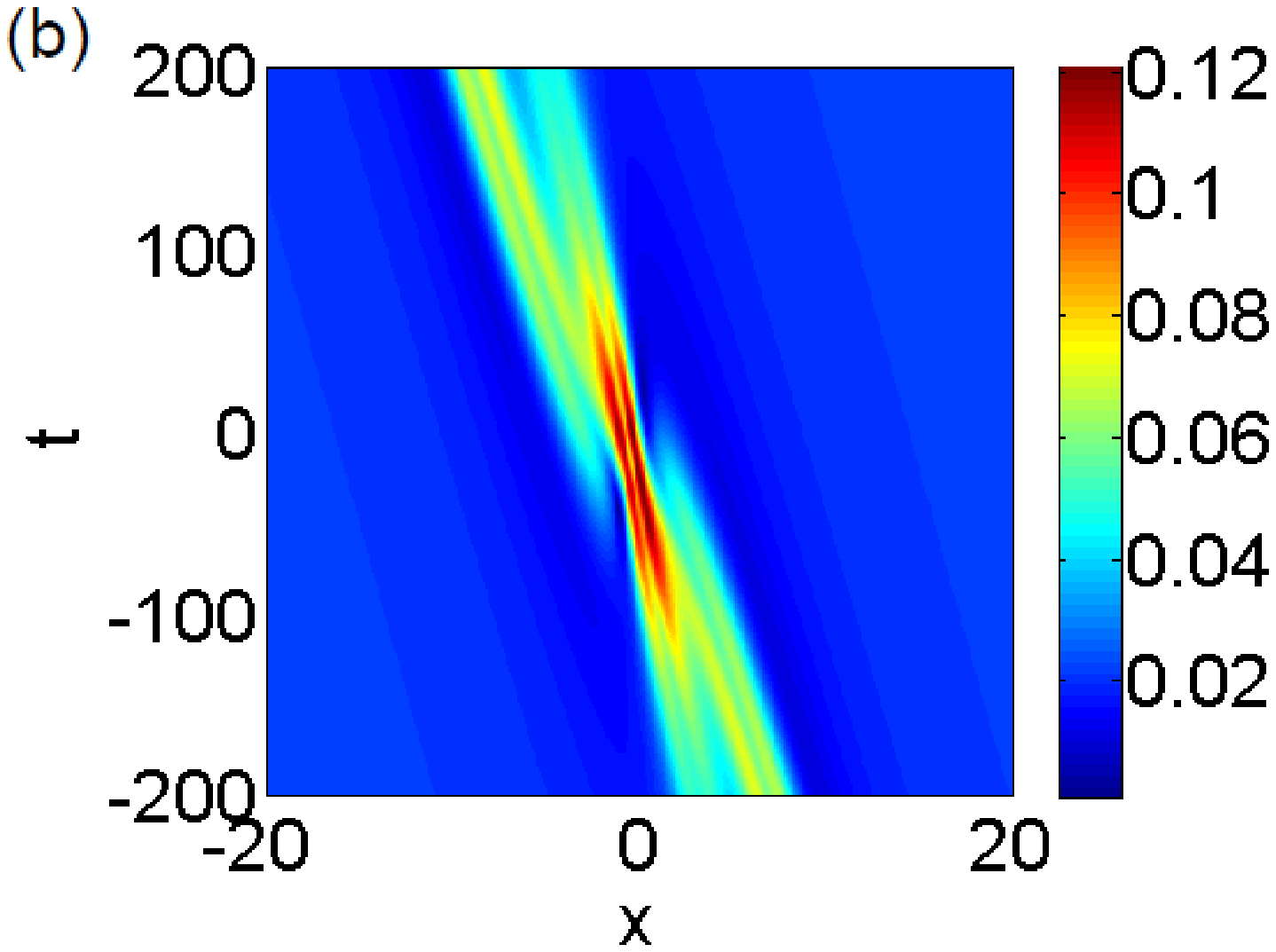}}
\caption{(a), (b) The second-order dark W-shaped soliton in the $E_1$ component and
double-peak W-shaped soliton the $p_2$ component, respectively.}
\label{fig:9}
\end{figure}

Suppose the discriminant of Eq. (\ref{61}) is zero under the zero frequency circumstance, one can have
\begin{equation}\label{62}
95 b_1^4-4580 b_1^3 b_2+15594 b_1^2 b_2^2-16388 b_1 b_2^3+1679 b_2^4=0.
\end{equation}
We point out that, when Eq. (\ref{62}) is satisfied, then the MI growth rate ${\rm Im}(\Omega)$ tends to zero.
At this time, rogue waves listed in Table 1
for each component of the CMB equations can transit into the corresponding W-shaped solitons.
Thus, solving the above equation and choosing
\begin{align}
&b_1=1, b_2=\dfrac{4097}{1679}+\dfrac{1328}{1679}\sqrt{6}-\dfrac{6}{1679}\sqrt{741210+303760\sqrt{6}},\nonumber\\
&m_1=-\dfrac{24715}{3358}-\dfrac{332}{1679}\sqrt{6}+\dfrac{3}{3358}\sqrt{741210+303760\sqrt{6}},\nonumber\\
&m_2=-\dfrac{20955}{3358}-\dfrac{2988}{1679}\sqrt{6}+\dfrac{27}{3358}\sqrt{741210+303760\sqrt{6}},\nonumber
\end{align}
we can obtain the first- and the second-order W-shaped soliton solutions by means of Eqs. (\ref{psi1}), (\ref{23}),
(\ref{r36}) and (\ref{r37}). Note that the state transition in each component of the CMB equations
can present different types of dynamic behaviors. For instance, the dark rogue wave
in the $E_1$ component can become a dark W-shaped soliton and the four-petaled rogue wave in the
$p_2$ component is changed into a double-peak W-shaped soliton,
see Figs. \ref{fig:8} and \ref{fig:9}.
Here we set $N_0=1$, $\gamma_0=0,\gamma_1=1,\gamma_2=0,\gamma_3=0$.

\section{Conclusion}

In conclusion, we investigated the CMB equations which govern the pulse propagation
in a medium with coherent three-level atomic transitions by virtue of the DT method.
Based on a $3\times 3$ linear eigenvalue problem with the AKNS form,
we constructed the $n$-fold DT in a compact determinant expression.
As an application, we derived the $n$th-order rogue wave solution
under the plane-wave solution (\ref{08}) for $a_2\neq a_1$ and $b_2\neq b_1$.
Dynamical behaviors of the $n$th-order rogue wave solution can be classified into two categories
by discussing the distribution of multiple root for the cubic spectral characteristic equation (\ref{09}), i.e.,
the double-root case and the triple-root case. With the help of the series expansions of the
eigenfunction expressed in powers of $\epsilon^2$ and $\epsilon^3$,
we obtained the explicit rogue wave solutions in terms of Schur polynomials.
For the double-root case, we showed the density plots of the novel
triple-hole rogue waves in the
$p_2$ component and the unusual twisted-pair rogue waves in the $M_{12}$ component
by choosing a set of specific parameters.
For the triple-root case, some novel composite rogue wave structures
were presented, for example, the composite four-petaled rogue waves in the $p_2$ component and the
composite dark rogue waves in the $M_{11}$ component.
On the other hand, we put forward the $n$th-order semirational solution
under the plane-wave solution (\ref{38}) for $a_2= a_1$ and $b_2= b_1$.
By expanding the eigenfunction in powers of $\epsilon^2$
on grounds of a double root of the quadratic equation (\ref{40}), we obtained the explicit
semirational solutions via the Schur polynomials.
Some novel interactional dynamics such as the interaction  between a dark soliton
and a dark rogue wave in the $N$ component and the
interaction between a bright soliton and a dark rogue wave in the $M_{11}$ component
were illustrated.
We further revealed the state transition between rogue wave and W-shaped soliton
on basis of the standard linearized stability analysis for the plane-wave solution (\ref{08}).
Analytical parametric condition related to the wavenumbers $b_1$ and $b_2$
under a group of specific parameters was obtained. The dark W-shaped solitons in the $E_1$ component
and the double-peak W-shaped solitons in the $p_2$ component were shown.
We hope the results obtained in this paper may help understand the intricate rogue wave
phenomena in areas of nonlinear optics, oceanography and so on.

\section*{Acknowledgment}
This work is supported by the National Natural Science Foundation of China
(Grant Nos. 11705290, 11705145, 11875126)
and the Young Scholar Foundation of ZUT (2018XQG16).


\end{document}